\documentclass[floats,preprint,prb,aps]{revtex4}
\begin{document}
 \preprint{{\em Submitted to Phys. Rev. {\bf A} } \hspace{3.7in} Web galley}
\title{Quasiparticle Density-Matrix Representation of Nonlinear TDDFT Response Functions}
\author{Oleg Berman$^{1}$ and Shaul Mukamel$^{1,2}$}
\affiliation{$^1$Department of Chemistry, University of Rochester,
Rochester, New York 14627-0216 \\
 $^2$Department of Physics and
Astronomy, University of Rochester, Rochester, NY 14627-0216}


\date{\today}

\vspace{2.7in}

\begin{abstract}
The time-dependent density functional (TDDFT) equations may be
written either for the Kohn-Sham orbitals in Hilbert space or for
the single electron density matrix in Liouville space. A
collective-oscillator, quasiparticle, representation of the
density response of many-electron systems which explicitly reveals
the relevant electronic coherence sizes is developed using the
Liouville space representation of adiabatic TDDFT. Closed
expressions for the nonlinear density-density response are
derived, eliminating the need to solve nonlinear integral
equations, as required in the Hilbert space formulation of the
response.

\vspace{1.0 in}



\end{abstract}

\maketitle {}

\newpage

\section{Introduction}

Time dependent density functional theory offers a computationally
tractable framework for calculating response functions of
many-electron systems (such as molecules, semiconductors and
metals) to electrical or optical perturbations
\cite{Hohenbergp:Inheg}$^-$\cite{Onida}. The linear and nonlinear
response functions (NRF) may then be used to extract useful
information about electronic excited states (energies, transition
and permanent dipole moments, etc.). The standard TDDFT algorithm
involves the computation of the Kohn-Sham orbitals
$\psi_{n}(\mathbf{r})$ in Hilbert space, \cite{Gross} and the
response functions are calculated in two steps: First, a reference
response is calculated for noninteracting electrons. Then a
nonlinear integral equation is solved for the actual response
function. The complexity of these calculations increases rapidly
with the nonlinear order of the response. Alternatively, the
Kohn-Sham equations may be written for the single electron density
matrix $\rho(\mathbf{r},\mathbf{r}')$ and its evolution in
Liouville space. We shall denote these as time dependent Hilbert
space (TDHS) and time dependent Liouville space (TDLS)
representations, respectively. The reduced single electron density
matrix for $N$ electron pairs $\psi_{n}(\mathbf{r},t)$ is given
by\cite{Coleman}
\begin{eqnarray} \label{rodef}
\rho (\mathbf{r}, \mathbf{r'},t) = \sum_{n=1}^{N}\psi
_{n}(\mathbf{r}, t)\psi _{n}^{*}(\mathbf{r'}, t)
\end{eqnarray}
A density matrix approach to the response was developed for the
time dependent Hartree-Fock approximation
(TDHF)\cite{Takahashia:Anhmns}$^{-}$\cite{Tretiak}, which is
formally very similar to TDDFT\cite{Chernyak2}. A closed algebra
of quasiparticles which satisfy a somewhat was established unusual
scalar product, which allows a systematic order by order expansion
of the density matrix  in the field \cite{Chernyak}. In this
article, we extend this approach to TDDFT and provide closed
expressions for density-density response functions using the
adiabatic exchange-correlation potential.

Doubling the space dimensionality in Liouville space compared to
Hilbert space $(\rho(\mathbf{r},\mathbf{r}')$ vs. $\psi_{n}
(\mathbf{r}))$ seems computationally prohibitive, in particular
since within the Kohn-Sham approximation
$\rho(\mathbf{r},\mathbf{r}')$ may be computed using $\psi_{n}
(\mathbf{r})$ using Eq.(\ref{rodef}). However, recasting reduced
descriptions of many-body systems using density
matrices\cite{Coleman,Mukamel} has many numerical advantages, as
will be discussed later. In addition, the TDLS representation
offers a valuable physical insight: (i) A quasiparticle picture of
the response is naturally obtained in terms of the eigenstates of
the linearized TDDFT. (ii) Closed expressions for response
functions may be derived. In contrast, in the TDHS approach one
needs to solve a chain of integral equations for each order
self-consistently. (iii) The density matrix carriers explicit
information about electronic coherence sizes. Molecular (e.g.
Kohn-Sham) orbitals can be arbitrarily transformed to be localized
or delocalized, without affecting the many-electron state. Their
coherence length is therefore ill defined and does not carry a
meaningful physical information. The density matrix is, on the
other hand, uniquely defined. Optical excitations involve the
creation of electron-hole pairs (excitons). The diagonal length of
$\rho (\mathbf{r},\mathbf{r}')$ along $\mathbf{r} = \mathbf{r}'$
reflects the center of mass motion of excitons whereas the
off-diagonal length (along $\mathbf{r}-\mathbf{r}'$) shows the
size associated with the relative electronic-hole pairs motion.
Plotting the density matrix provides physical insights and offers
a real space N scaling description of electronic excitations when
the density matrix is localized along $\mathbf{r}-\mathbf{r}'$,
which is typically the case~\cite{White}. iv) The response
functions for any single-electron operator (e.g. current) can be
readily obtained. This allows to compute, for example,
conductivities.

This paper is organized as follows: The time dependent Kohn-Sham
equations of motion for the single electron density matrix are
presented in Sec. II. The CEO (collective electron oscillator),
quasiparticle, representation of the TDDFT density matrix is
derived in Sec.III. The linear density-density response function
is calculated in Sec.IV and the second-order and the third-order
responses are given in Appendixes B and C, respectively. The
advantages of the TDLS approach for computations of linear and
non-linear time-dependent response functions, as well as possible
extensions are summarized in Sec.V.

\section{Time dependent Kohn-Sham equations for the
single electron density matrix} \label{sec.ham1}

The time dependent Kohn-Sham equations of motion  for the density
matrix~\cite{Chernyak2} are
\begin{eqnarray} \label{eqro}
  i \frac{\partial \rho (\mathbf{r}, \mathbf{r'},t)}{\partial t} =
  \widehat L_{KS}[\rho (\mathbf{r}, \mathbf{r'},t)](\mathbf{r}, \mathbf{r'},t) \rho (\mathbf{r}, \mathbf{r'},t) ;
\end{eqnarray}
where we set ($\hbar = 1$) and
\begin{eqnarray} \label{ro}
\rho(\mathbf{r},t) = \rho(\mathbf{r},\mathbf{r},t ) \equiv
n(\mathbf{r},t)
\end{eqnarray}
is the exact time-dependent charge density of electrons; $\widehat
L_{KS}[\rho (\mathbf{r}, \mathbf{r'},t)](\mathbf{r},
\mathbf{r'},t)$ is the superoperator in Liouville space
corresponding to the Kohn-Sham Hamiltonian.

\begin{eqnarray} \label{lks}
   \widehat L_{KS}[\rho (\mathbf{r}, \mathbf{r'},t)](\mathbf{r}, \mathbf{r'},t) \equiv
    \widehat H_{KS}[\rho (\mathbf{r},\mathbf{r},t)](\mathbf{r}, t) - \widehat H_{KS}^{*}[\rho (\mathbf{r'},\mathbf{r'},t)](\mathbf{r'}, t);
\end{eqnarray}
$\widehat H_{KS}$ is the Kohn-Sham Hamiltonian
\begin{eqnarray} \label{hks0}
 \widehat H_{KS}= \widehat H_{KS}^{0}  + U_{1}(\mathbf{r}, t) .
\end{eqnarray}
$H_{KS}^{0}$ is the Hamiltonian of the isolated model and $U_{1}$
represents the coupling to an external optical field.
\begin{eqnarray} \label{hks}
   \widehat H_{KS}^{0}[\rho (\mathbf{r}, \mathbf{r},t)](\mathbf{r},t) =
    -\frac{\nabla_{{\mathbf{r}}}^{2}}{2m} + U_{KS}^{0}[\rho (\mathbf{r},\mathbf{r},t)](\mathbf{r}) + U_{0}(\mathbf{r});
\end{eqnarray}
$m$ is an effective  electron mass and  the Kohn-Sham external
potential is a functional of the charge density:
\begin{eqnarray} \label{uks0}
  U_{KS}^{0}[\rho (\mathbf{r},\mathbf{r},t)](\mathbf{r}) = \int d\mathbf{r'}\frac{\rho
  (\mathbf{r'},\mathbf{r'},t)}{|\mathbf{r}-\mathbf{r'}|} + U_{xc}[\rho
(\mathbf{r},\mathbf{r},t)](\mathbf{r}) ;
\end{eqnarray}
$U_{0}(\mathbf{r})$ is the field created by nuclei and
$U_{xc}[\rho (\mathbf{r},\mathbf{r},t)](\mathbf{r})$ is the
exchange-correlation potential in the adiabatic approximation. We
further define the time-dependent external potential, given by
$U_{ext}(\mathbf{r}, t) = U_{0}(\mathbf{r})$ at time $t\leq t_{0}$
and $U_{ext}(\mathbf{r}, t) = U_{0}(\mathbf{r})+ U_{1}(\mathbf{r},
t)$ for $t > t_{0}$.

\section{Quasiparticle representation of density response functions} \label{sec.ham}

The TDDFT equations describe the evolution of the reduced
single-electron density matrix of a many-electron system driven by
an external field. The calculation starts by computing the ground
state single electron density matrix $\bar{\rho}
(\mathbf{r},\mathbf{r'})$ in real space, which is determined by
the stationary solution of the Kohn-Sham equation
(Eq.~(\ref{eqro})). $\bar{\rho}$ is a key input to the TDDFT
calculations. We then set $\rho(\mathbf{r},\mathbf{r'},t)=\bar
\rho (\mathbf{r},\mathbf{r'}) +
\delta\rho(\mathbf{r},\mathbf{r'},t)$ where $\delta \rho
(\mathbf{r},\mathbf{r'},t)$ represents the changes induced by the
external field, described by the potential $U_{1}(\mathbf{r}, t)$.
The diagonal elements $\delta \rho (\mathbf{r},\mathbf{r},t)$ give
the changes in charge density, whereas the off-diagonal elements $
\rho (\mathbf{r},\mathbf{r'},t)$ represent the changes in
electronic coherences between two points. Eq.(\ref{eqro}) is then
recast in the form
 \begin{eqnarray}
  i\frac{\partial \delta \rho(\mathbf{r},\mathbf{r'},t)}{\partial t}
   &=& ( \widehat H_{KS}^{0}[\rho (\mathbf{r},\mathbf{r},t)](\mathbf{r}) - \widehat H_{KS}^{0*}[\rho (\mathbf{r'},\mathbf{r'},t)](\mathbf{r'})    \nonumber \\
   &+& [U_{1}(\mathbf{r}, t) - U_{1}^{*}(\mathbf{r'}, t)] ) \rho(\mathbf{r},\mathbf{r'},t);
  \label{eqtdhfrho}
 \end{eqnarray}
This equation may be solved for $\delta
\rho(\mathbf{r},\mathbf{r'},t)$ either in the frequency or in the
time domain.

In order to calculate linear and nonlinear response functions we
expand $\delta\rho(\mathbf{r},\mathbf{r'},t)$ in powers of $U
_{1}(\mathbf{r}, t)$: $\delta\rho(\mathbf{r},\mathbf{r'},t) =
\delta \rho _{1} (\mathbf{r},\mathbf{r'},t) + \delta\rho
_{2}(\mathbf{r},\mathbf{r'},t) + \delta\rho
_{3}(\mathbf{r},\mathbf{r'},t) + ... $ and solve the resulting
equations of motion successively order by order. To that end we
first expand $U_{KS}^{0}$ in Taylor series in the vicinity of
$\bar \rho$ to third order in $\delta \rho$:
\begin{eqnarray} \label{taylor}
 && U_{KS}^{0}[\rho(\mathbf{r},t)](\mathbf{r}) = U_{KS}^{0}[\bar \rho](\mathbf{r}) + V_{f} + V_{g} + V_{h};  \nonumber \\
 && V_{f} = \int  d\mathbf{r}' \left( \frac{e^2}{|\mathbf r - \mathbf r'|} + f_{xc}[\bar \rho] (\mathbf r, \mathbf r') \right)
 \delta \rho (\mathbf{r}',t) ; \nonumber \\
 && V_{g} =  \int d\mathbf{r}' \int d\mathbf{r}'' g_{xc}[\bar \rho] (\mathbf r, \mathbf r', \mathbf{r}'')
 \delta \rho (\mathbf{r}',t) \delta \rho (\mathbf{r}'',t); \nonumber \\
 && V_{h} =  \int d\mathbf{r}' \int d\mathbf{r}'' \int d\mathbf{r}''' h_{xc}[\bar \rho] (\mathbf r, \mathbf r', \mathbf{r}'',\mathbf{r}''')
 \delta \rho (\mathbf{r}',t) \delta \rho (\mathbf{r}'',t) \delta \rho
 (\mathbf{r}''',t) ,
\end{eqnarray}
where for brevity we denote $\delta\rho(\mathbf{r},\mathbf{r},t)$
as $\delta\rho(\mathbf{r},t)$; $e$ is the electron charge and
$f_{xc}[\bar \rho] (\mathbf r, \mathbf r')$, $g_{xc}[\bar \rho]
(\mathbf r, \mathbf r', \mathbf r'')$ and $h_{xc}[\bar \rho]
(\mathbf r, \mathbf r', \mathbf r'', \mathbf r''')$ are the first,
the second and the third order adiabatic exchange-correlation
kernels \cite{Gross} (we consider the commonly used adiabatic
approximation where they are assumed time-independent):
\begin{eqnarray} \label{fxct1}
f_{xc}[\bar \rho] (\mathbf r, \mathbf r') = \left. \frac{\delta
U_{xc}[\rho](\mathbf{r})}{\delta \rho(\mathbf{r'},t')}
\right|_{\bar \rho},
\end{eqnarray}
\begin{eqnarray} \label{gxct1}
g_{xc}[\bar \rho] (\mathbf r, \mathbf r',\mathbf r'') = \left.
\frac{\delta^{2} U_{xc}[\rho](\mathbf{r})}{\delta
\rho(\mathbf{r'},t') \delta \rho(\mathbf{r''},t'')} \right|_{\bar
\rho},
\end{eqnarray}
\begin{eqnarray} \label{hxct1}
h_{xc}[\bar \rho] (\mathbf r, \mathbf r',\mathbf r'',\mathbf r''')
= \left. \frac{\delta^{3} U_{xc}[\rho](\mathbf{r})}{\delta
\rho(\mathbf{r'},t') \delta \rho(\mathbf{r''},t'')\delta
\rho(\mathbf{r'''},t''')} \right|_{\bar \rho}.
\end{eqnarray}
We further define
\begin{eqnarray} \label{fxct11}
f_{xc}'[\bar \rho] (\mathbf r, \mathbf r') \equiv f_{xc}[\bar
\rho] (\mathbf r, \mathbf r') + \frac{e^2}{|\mathbf r - \mathbf
r'|}
\end{eqnarray}
and introduce the following notation for the product of two real
space matrices $\zeta$ and $\eta$
\begin{eqnarray}
\label{star} (\zeta * \eta )(\mathbf{r},\mathbf{r}') \equiv \int
\zeta (\mathbf{r},\mathbf{r}'') \eta (\mathbf{r}'',\mathbf{r}')
d\mathbf{r}'' .
\end{eqnarray}
For an operator $A[\zeta (\mathbf{r},\mathbf{r})](\mathbf{r})$ and
a matrix $\eta (\mathbf{r},\mathbf{r'})$ we denote
 \begin{equation}
 [A,\eta](\mathbf{r},\mathbf{r'}) \equiv \left( A[\zeta (\mathbf{r},\mathbf{r})](\mathbf{r}) - A^{*}[\zeta(\mathbf{r'},\mathbf{r'})](\mathbf{r'})\right) \eta(\mathbf{r},\mathbf{r'}) ,
  \label{oper}
 \end{equation}
and the commutator of two single-electron matrixes $\zeta$ and
$\eta$
 \begin{equation}
 [\zeta,\eta](\mathbf{r},\mathbf{r'}) \equiv (\zeta * \eta )(\mathbf{r},\mathbf{r'}) - (\eta * \zeta
 )(\mathbf{r},\mathbf{r'}).
 \label{matr}
 \end{equation}

Since $\rho (\mathbf{r},\mathbf{r'},t)$ represents a single Slater
determinate, it must be idempotent, and, consequently, not all
elements of $\delta \rho (\mathbf{r},\mathbf{r'},t)$ are
independent\cite{Chernyak}. When expanded using the ground state
 Kohn-Sham orbitals, $\delta\rho$ can be divided into four
blocks: electron-hole and hole-electron (interband) and
electron-electron and hole-hole (intraband). Only the interband
elements of $\xi (\mathbf{r},\mathbf{r'},t)$ are independent and
the intraband blocks $T$ can be expressed in terms of these
elements~\cite{thoulessbook,ring,blaizot}. We thus decompose
$\delta \rho(\mathbf{r},\mathbf{r'},t)$ as (see Appendix A)
 \begin{equation}\label{decompos}
  \delta \rho(\mathbf{r},\mathbf{r'},t) \equiv \xi(\mathbf{r},\mathbf{r'},t) + T(\xi(\mathbf{r},\mathbf{r'},t)),
 \end{equation}
It will be desirable to remain in real space and avoid switching
back and forth to the orbital basis. Fortunately, this goal can be
accomplished and $\xi$ can be computed directly in real space
using the double commutator:
\begin{equation}\label{projectxi}
\xi(\mathbf{r},\mathbf{r}',t) \equiv [[\delta\rho (t),\bar
\rho],\bar \rho] (\mathbf{r},\mathbf{r}') .
\end{equation}
The operation in the r.h.s. is a {\it projection operator} that
projects $\delta \rho$ into the interband  subspace (see
Eq.(\ref{project}))\cite{Chernyak}. Once $\xi$ is calculated we
 obtain for $T$ (see Appendix A)
\begin{eqnarray}
T(\xi (\mathbf{r},\mathbf{r'},t))= (I-2\bar \rho )*(\xi(t)*\xi(t)
+\xi(t)*\xi(t)*\xi(t)*\xi(t) +  \cdots)(\mathbf{r},\mathbf{r'}),
\label{EQTO}
\end{eqnarray}
where all $\xi(\mathbf{r},\mathbf{r'},t)$ are taken at time t.

The next step is to develop a convenient quasiparticle algebra for
the independent dynamical variables $\xi$ which represent
collective electronic excitations. To that end we start with the
eigenvalue equation corresponding to the linearized Kohn-Sham
equations
\begin{eqnarray}
\label{eigenvalue}
 L \xi_\alpha (\mathbf r, \mathbf r') =
\Omega_\alpha \xi_\alpha (\mathbf r, \mathbf r') ,
\end{eqnarray}
where
 \begin{eqnarray}
 \label{Lf}
L = H_{f}(\mathbf{r})- H_{f}^{*}(\mathbf{r}')  ,
  \end{eqnarray}
 \begin{eqnarray}
 \label{Hf}
H_{f}(\mathbf{r}) \xi_\alpha (\mathbf r, \mathbf r') =
-\frac{\nabla_{\mathbf{r}}^{2}\xi_\alpha (\mathbf r, \mathbf
r')}{2m} + \left(\int d\mathbf{r}' \int d\mathbf{r}'' f_{xc}'[\bar
\rho] (\mathbf r, \mathbf r', \mathbf r'')\xi_\alpha (\mathbf r',
\mathbf r'') \right) \bar \rho (\mathbf{r},\mathbf{r}') ,
  \end{eqnarray}
and
 \begin{eqnarray}
 \label{f''}
f_{xc}'[\bar \rho] (\mathbf r, \mathbf r', \mathbf r'') =
f_{xc}'[\bar \rho] (\mathbf r, \mathbf r')\delta(\mathbf r'-
\mathbf r'').
  \end{eqnarray}
The eigenmodes $\xi_\alpha$ come in pairs. Each pair of conjugated
modes $\xi_{\alpha}(\mathbf r, \mathbf r')$ and
$\xi_{\alpha}^\dagger(\mathbf r, \mathbf r')$ forms a collective
electronic oscillator $\alpha$ (where $\alpha = 1, 2, \ldots $).
We define $\xi_{-\alpha}(\mathbf r, \mathbf
r')=\xi_{\alpha}^\dagger(\mathbf r, \mathbf r')$ and
$\Omega_{-\alpha}=-\Omega_{\alpha}$ where $\alpha = \pm 1, \pm 2,
\ldots $. The algebraic properties of these oscillators were
developed in Ref.\cite{Chernyak} and reviewed in
Ref.\cite{Tretiak}.

We further introduce the following scalar product of any two
interband matrices $\xi$ and $\eta$.
\begin{equation}
\label{scalar}
 \langle \xi|\eta \rangle \equiv \int d\mathbf{r} \int d\mathbf{r'}\bar
\rho[\xi^\dagger,\eta](\mathbf{r},\mathbf{r}')\delta(\mathbf{r} -
\mathbf{r}').
 \end{equation}
The bra (ket) notation underscores the similarity with Dirac's
Hilbert space notation. Eq.~(\ref{scalar}) satisfies the following
relations:
\begin{eqnarray}
\label{scalarpr}
 \langle \xi|\eta \rangle =\langle \eta^\dagger
|\xi^\dagger \rangle^*=-\langle \eta|\xi \rangle\ .
\end{eqnarray}
Using this scalar product, we shall normalize these modes as
follows
 \begin{equation}
 \langle \xi_\alpha^\dagger |\xi_\beta \rangle = \delta_{\alpha \beta},
 \hspace{5em}
 \langle \xi_\alpha|\xi_\beta \rangle =  0 .
 \label{norm1}
 \end{equation}
Eqs.~(\ref{scalarpr}) and~(\ref{norm1}) show that this is an
unusual scalar product: for one thing, the norm of a vector is
zero. Nevertheless, it has one important property: it allows us to
expand $\xi(\mathbf r, \mathbf r', t)$ in terms of collective
electron oscillator (CEO) modes $\xi_{\alpha}(\mathbf r, \mathbf
r')$
\begin{eqnarray}  \label{eqosc6}
\xi (\mathbf r, \mathbf r', t)= \sum_{\alpha = \pm 1, \pm2 \ldots
}  \xi_{\alpha}(\mathbf r, \mathbf r') z_{\alpha}(t),
\end{eqnarray}
where $z_{\alpha}(t) = \langle \xi_\alpha^\dagger |\xi (t)\rangle$
and its complex conjugate $z_{-\alpha}(t) =z_{\alpha}^*(t)$
constitute complex oscillator amplitudes. $\xi_{\alpha}$ thus form
a complete basis for representing an arbitrary interband matrix.

Combining Eq.(\ref{eqosc6}) and Eq.(\ref{decompos}), we get
\begin{eqnarray} \label{eqpolar2}
\delta \rho (\mathbf r,\mathbf r', t) &=& \sum_{\alpha}
\rho_{\alpha}(\mathbf r,\mathbf r') z_{\alpha}(t) + \frac {1}{2}
\sum_{\alpha \beta } \rho_{\alpha \beta}(\mathbf r,\mathbf r')
z_{\alpha}(t) z_{\beta}(t) \nonumber \\
&+& \frac {1}{3} \sum_{\alpha \beta \gamma} \rho_{\alpha, \beta
\gamma}(\mathbf r,\mathbf r')z_{\alpha}(t)
z_{\beta}(t)z_{\gamma}(t), \nonumber \\ && \alpha,  \beta, \gamma
= \pm 1, \pm 2, \ldots
\end{eqnarray}
where we only retained terms that contribute to the third order
response. The expansion coefficients in the r.h.s. are given by
\begin{eqnarray} \label{3.0a}
\rho_{\alpha}(\mathbf r,\mathbf r') = \xi_{\alpha}(\mathbf r,
\mathbf r'),
\end{eqnarray}
\begin{eqnarray} \label{3.0aa}
 \rho_{\alpha \beta} (\mathbf r,\mathbf r') =  (I - 2 \bar \rho)*(\xi_{\alpha}* \xi_{\beta} + \xi_{\beta} * \xi_{\alpha})(\mathbf r, \mathbf r'),
\end{eqnarray}
\begin{eqnarray} \label{3.0aaa}
 \rho_{\alpha, \beta\gamma} (\mathbf r,\mathbf r') = - \xi_{\alpha}*(\xi_{\beta}*\xi_{\gamma} + \xi_{\gamma}*\xi_{\beta})(\mathbf r, \mathbf r').
\end{eqnarray}
The amplitudes of the adjoint (negative frequency) variables are
simply the complex conjugates ($z_{-\alpha} = z_{\alpha}^{*}$).

Upon the substitution of
Eqs.~(\ref{eqpolar2}),(\ref{3.0a}),(\ref{3.0aa}), and
(\ref{3.0aaa}) into Eq.~(\ref{eqtdhfrho}) we obtain the equation
of motion for the CEO amplitude $z(t)$ to third order in the
external field:
\begin{eqnarray} \label{eqtdhf22222}
i \frac{\partial z_{\alpha}(t) }{\partial t} &=& \Omega_{\alpha}
z_{\alpha} (t) + K_{-\alpha} +  \sum_{\beta}K_{-\alpha \beta} \
z_{\beta}(t) + \sum_{\beta\gamma}K_{-\alpha \beta \gamma} \
z_{\beta}(t)z_{\gamma}(t) + \sum_{\beta \gamma
\delta}K_{-\alpha\beta\gamma\delta} \ z_{\beta}(t)
z_{\gamma}(t)z_{\delta}(t); \nonumber \\
&& \hspace{0.5em}
 \alpha, \beta,\gamma,\delta = \pm 1, \pm 2, \ldots,
\end{eqnarray}
where
\begin{eqnarray} \label{eqtdhf22333}
&& K_{-\alpha} = \int U_{1}(\mathbf r , t) \rho_{-\alpha}(\mathbf
r) d\mathbf{r} ;  \hspace{1.5em} \rho_{-\alpha}(\mathbf r) \equiv
\rho_{-\alpha}(\mathbf r,\mathbf
r ); \nonumber \\
&& K_{-\alpha \beta} = \int U _{1}(\mathbf r, t)
\rho_{-\alpha,\beta}(\mathbf r) d\mathbf{r} ; \hspace{1.5em}
\rho_{-\alpha,\beta}(\mathbf r) \equiv
\rho_{-\alpha,\beta}(\mathbf r,\mathbf
r ); \nonumber \\
&& K_{-\alpha \beta\gamma} = U_{KS(-\alpha\beta\gamma)}^{0} + \int
U _{1}(\mathbf r, t) \rho_{-\alpha,\beta\gamma}(\mathbf r)
d\mathbf{r} ;  \hspace{1.5em} \rho_{-\alpha,\beta\gamma}(\mathbf
r) \equiv \rho_{-\alpha,\beta\gamma}(\mathbf r,\mathbf
r ); \nonumber \\
&& K_{-\alpha \beta\gamma\delta} =
U_{KS(-\alpha\beta\gamma\delta)}^{0},
\end{eqnarray}
$U_{KS}^{0}$ is the Kohn-Sham external potential, expanded to the
third-order kernel Eq.~(\ref{taylor}), and
$U_{KS(-\alpha\beta\gamma)}^{0}$ and
$U_{KS(-\alpha\beta\gamma\delta)}^{0}$ are given in
Eq.~(\ref{vabg}) and Eq.~(\ref{vabgd}), respectively.

Eqs.~(\ref{eqtdhf22222}) together with~(\ref{eqpolar2}) constitute
the quasiparticle algorithm for computing density response
functions. These nonlinear equations which map the system onto
coupled collective classical oscillators, may be solved by
expanding $z(t)$ ($z^*(t)$) in powers of the external field ${U
_{1}}(\mathbf r, t)$:
\begin{equation} \label{zexpan}
 z(t)= z^{(1)}(t)+z^{(2)}(t)+ z^{(3)}(t) + \ldots,
\end{equation}
Substituting this in Eq.~(\ref{eqtdhf22222}) we can successively
solve for $z^{(j)}$ order by order which when substituted in
Eq~(\ref{eqpolar2}) will give $\delta\rho_{j}$.  The present
truncation of Eqs.~(\ref{eqpolar2}) and~(\ref{eqtdhf22222}) allows
to compute response functions up to third order. However, this
approach may be extended to NRF of arbitrary order by simply
adding higher terms to Eqs.~(\ref{eqpolar2})
and~(\ref{eqtdhf22222}).

\section{The linear density response}

The following calculation of the linear response illustrates the
strategy for computing response functions. The first-order density
matrix (Eq.~(\ref{eqtdhfrho})) satisfies:
 \begin{eqnarray}
 \label{rho1}
  i\frac{\partial \delta \rho_{1}(\mathbf{r},\mathbf{r}',t)}{\partial t}
  = L \delta \rho_{1}(\mathbf{r},\mathbf{r}',t) + [U_{1}(\mathbf{r}, t) - U_{1}^{*}(\mathbf{r'}, t)]\delta \rho_{1}(\mathbf{r},\mathbf{r'},t).
  \end{eqnarray}
The equation of motion for $z_{\alpha}^{(1)}(t)$ is obtained from
Eq.~(\ref{rho1}) together with Eq.(\ref{decompos}) and the
expansion Eq.~(\ref{EQTO}):
\begin{eqnarray} \label{zlin}
i \frac {\partial z_{\alpha}^{(1)}(t)}{\partial t} &=&
\Omega_{\alpha} z_{\alpha}^{(1)}(t) + \int  {U _{1}}(\mathbf r ,
t) \rho_{-\alpha}(\mathbf r) d\mathbf{r} , \hspace{1em} \alpha =
\pm 1, \pm 2, \ldots
\end{eqnarray}
The solution of this equation can be represented as
\begin{eqnarray} \label{zlinsol3}
z_{\alpha}^{(1)}(t) & = & i s_\alpha \int_{-\infty}^t d\tau  \int
d\mathbf r {U _{1}} (\mathbf r,\tau) \rho_{-\alpha}(\mathbf r)
G_{\alpha} (t - \tau),
\end{eqnarray}
where positive and negative $\alpha$ correspond to
$z_{\alpha}^{(1)}(t)$ and ${z_{\alpha}^*}^{(1)}(t)$, respectively,
$s_{\alpha} \equiv sign(\alpha)$, and we have introduced the
time-domain Green function
\begin{equation}\label{green}
  G_\alpha(t) = \theta(t) e^{-i\Omega_\alpha t},  \hspace{3em}
  G_{-\alpha}(t) = \theta(t) e^{-i\Omega_{-\alpha} t}=\theta(t) e^{i\Omega_\alpha t},
\end{equation}
where $\theta(t)$ is the Heavyside step function.

Inserting Eq.~(\ref{zlinsol3}) into Eq.~(\ref{eqpolar2}) we obtain
for the density to linear order
\begin{eqnarray} \label{pollin}
\delta\rho_{1} (\mathbf r, t) &=& \delta\rho_{1} (\mathbf r,
\mathbf r, t) = \sum_{\alpha=\pm 1, \pm2, \ldots}
z^{(1)}_{\alpha}(t) \rho_{\alpha}(\mathbf r).
\end{eqnarray}
This can be recast in the form
\begin{eqnarray} \label{rest1}
\delta\rho_{1} (\mathbf r, t) &=& \int_{-\infty}^t d \tau \int
d\mathbf{r'} {U _{1}}(\mathbf r',\tau) \chi^{(1)} (t, \tau,\mathbf
r, \mathbf r'),
\end{eqnarray}
where the first order time-domain linear density-density response
function is given by \cite{Gross}:
\begin{eqnarray} \label{rest1a}
\chi^{(1)} (t, \tau, \mathbf r, \mathbf r') &=&  i \sum_{\alpha=
\pm 1, \pm2, \ldots} i s_\alpha \rho_{-\alpha}(\mathbf r')
\rho_{\alpha}(\mathbf r) G_{\alpha} (t - \tau).
\end{eqnarray}

The corresponding frequency-domain density-density linear response
function $\chi^{(1)}(-\omega_s; \omega)$ is defined by
\begin{eqnarray} \label{resf1}
\delta\rho_{1} (\mathbf r, \omega_s) &=& \int_{-\infty}^{\infty}
\frac {d \omega}{2 \pi}\int d\mathbf{r'} \chi^{(1)} (-\omega_s;
\omega,\mathbf r, \mathbf r') {U _{1}} (\mathbf r',\omega).
\end{eqnarray}
Here ${U _{1}} (\mathbf r, \omega)$ is the Fourier transform of
the time-dependent external field ${U _{1}} (\mathbf r, t)$
defined as
\begin{equation}\label{fourier}
  f(\mathbf r, \omega) \equiv \int dt f(\mathbf r, t) e^{i\omega t}\;, \hspace{2em}
  f(\mathbf r, t) \equiv \frac{1}{2 \pi} \int d\omega f(\mathbf r,\omega) e^{-i\omega t}.
\end{equation}
By comparing Eqs.~(\ref{rest1}) with Eqs.~(\ref{resf1}) and using
Eq.~(\ref{fourier})\cite{Mukamel,Bloembergen} we have:
\begin{eqnarray} \label{restf1}
&&\chi^{(1)} (-\omega_s; \omega,\mathbf r, \mathbf r') =
\int_{-\infty}^{\infty} d t e^{i \omega_s t} \int_{-\infty}^t d
\tau  e^{-i \omega \tau}\chi^{(1)} (t, \tau,\mathbf r, \mathbf
r'),
\end{eqnarray}
The linear order response functions is usually denoted
\cite{Bloembergen}
\begin{eqnarray} \label{restf1a}
\chi^{(1)}(\omega_s=\omega; \omega, \mathbf r, \mathbf r') & = &
2\pi\delta(-\omega_s+\omega)\chi^{(1)} (\omega, \mathbf r, \mathbf
r'),
\end{eqnarray}
Using Eqs.~(\ref{restf1}) and (\ref{restf1a}) we obtain the linear
density-density response function
\begin{eqnarray} \label{pol1}
\chi^{(1)}(\omega,\mathbf r, \mathbf r') & = & \sum_{\alpha = \pm
1, \pm2, \ldots} \frac{s_{\alpha} \rho_{-\alpha}(\mathbf r')
\rho_{\alpha}(\mathbf r)}{\Omega_{\alpha} - \omega} = \sum_{\alpha
= 1, 2, \ldots} \frac{2 \Omega_\alpha \rho_\alpha (\mathbf
r)\rho_\alpha (\mathbf r')} {\Omega_\alpha^2 - \omega^2}.
\end{eqnarray}
Here and below $\Omega_\alpha$ is positive (negative) for all
$\alpha>0$ ($\alpha<0$), following the convention
$\Omega_{-\alpha}=-\Omega_\alpha$.

Finally, the static linear density-density response can be
obtained from Eq.~(\ref{pol1}) by setting $\omega=0$:
\begin{eqnarray} \label{pol1a}
\chi^{(1)} (0,\mathbf r, \mathbf r') & = & \sum_{\alpha = 1, 2,
\ldots} \frac{2 \rho_\alpha (\mathbf r)\rho_\alpha (\mathbf r')}
{\Omega_\alpha}.
\end{eqnarray}

Higher order response functions may be computed in the same way
and depend on the same ingredients: the collective electronic
oscillator (CEO) modes. Closed expressions for the second-order
density-density response functions are given in Appendix B
(Eqs.~(\ref{resp2}),~(\ref{pol2}) and~(\ref{pol2a})) and the third
order response is presented in Appendix C
(Eqs.~(\ref{resp3}),~(\ref{pol3}) and~(\ref{pol3a})).

\section{Discussion} \label{ap.4thdiscussion}

The density matrix TDLS representation of linear and non-linear
TDDFT density-density response functions has several notable
computational advantages compared with the Hilbert space (orbital)
TDHS representation. To facilitate the comparison, we have
outlined the TDHS in Appendix D. In the TDLS approach we need to
solve the CEO eigenfunction and eigenvalue problem in Liouville
space (Eq.~(\ref{eigenvalue})) to obtain the quasiparticle
spectrum of excitations (the CEO modes) \cite{Chernyak}.
Eqs.(\ref{pol1}), (\ref{pol2}) and (\ref{pol3}) can then be
directly used to  calculate time-dependent response functions in a
single step. In TDHS, in contrast, we proceed in two steps: we
first solve $K$ Kohn-Sham equations Eq.(\ref{eqphi}) with
Eq.(\ref{n}) for the Kohn-Sham orbitals, and calculate the linear
response for non-interacting particles Eq.(\ref{xis}) and as a
second step, we solve the Dyson-type integral equation
Eq.(\ref{xi1int}) to obtain linear response \cite{Gross}.
Calculating the non-linear response in this approach is even more
complicated. For example, obtaining the second-order response
requires solving the Dyson-type integral equation
Eq.(\ref{xi2int}) using the integral equation for the linear
response Eq.(\ref{n1int}). Similarly, obtaining the third-order
response requires the solution of Dyson-type integral equation
Eq.(\ref{n3int}) using the integral equations for the second
Eq.(\ref{n2int}) and linear response Eq.(\ref{n1int}). The
computation of non-linear higher-order response thus involves the
chain of self-consistent integral equations for all previous
responses, while the TDLS provides a closed expression for the
response functions. Clearly, the TDLS approach provides a much
faster one-step algorithm for computations of response functions.

The CEO modes are nonlocal and act in a six dimensional space
$(\mathbf r, \mathbf r')$ compared to the three dimensional space
$(\mathbf r)$ of the Kohn-Sham orbitals. However, several points
help reduce the computational cost. First, we can reduce the
number of Liouville space equations by solving only for
particle-hole (interband) density matrix $\xi$ instead of the
entire density matrix $\delta \rho$: the particle-particle and
hole-hole (intraband) density matrix $T(\xi)$ can be obtained from
Eq.~(\ref{EQTO}), which follows from the idempotent property of
single Slater determinant. For a basis set of $K$ orbitals the
number of elements is the $KN$ rather than $K^2$.  Moreover, the
density matrices have non-vanishing elements only when
$|\mathbf{r} - \mathbf{r}'|$ is less than a coherence size, which
is typically very short. This allows to neglect many density
matrix elements, further reducing the size. It is not possible to
include the coherent size in TDHS computations.

The present work can be extended in several ways. Even though we
have derived closed expressions for the response up to third
order, the computation of higher-order responses is
straightforward and merely involves carrying out the expansions in
Eqs.~(\ref{eqpolar2}) and~(\ref{eqtdhf22222}). The procedure, used
here to compute response functions of the charge density, can be
used to obtain response functions for any single-electron
operator. For example, conductivities are given by the current
response
\begin{equation}
  j^{(i)}(\mathbf r,t)= Tr((v \delta \rho^{(i)}(t))(\mathbf r, \mathbf r')),
  \label{chij}
\end{equation}
where $v(\mathbf r,\mathbf r') = -\frac{1}{2m}(\nabla_{\mathbf r'}
- \nabla_{\mathbf r}) $ is the velocity operator.

Another interesting possible extension of this work is to include
non-adiabatic exchange-correlation potentials, as outlined
recently for the linear response~\cite{Chernyak2}. In general, the
exchange-correlation potential and exchange-correlation kernels
are time-dependent~\cite{GrossKohn}. This time dependence has been
neglected within the adiabatic approximation used here. If we
relax this approximation, the eigenvalue equation for the
Liouville superoperator $L$, Eq.~(\ref{eigenvalue}), should be
replaced by~\cite{Chernyak2}
\begin{eqnarray}
\label{eigenvalue2}
 L(\Omega_{\alpha}) \xi_\alpha (\mathbf r, \mathbf r') =
\Omega_\alpha \xi_\alpha (\mathbf r, \mathbf r') .
\end{eqnarray}
Methods for solving Eq.(\ref{eigenvalue2}) using the
frequency-dependent functional of Gross and Kohn~\cite{GrossKohn}
were described in~\cite{Chernyak2}.

The time dependent Hartree-Fock TDHF (also known as RPA) is an
alternative widely used approximation for computing response
functions \cite{thoulessbook,ring,blaizot}. Within the density
matrix representation the TDHF is formally equivalent to TDDFT,
provided we use the following exchange-correlation potential that
depends on the density matrix (rather than merely on the charge
density) \cite{Chernyak2}:
\begin{eqnarray} \label{tdhfexcorr}
U_{KS-TDHF}^{0}[\rho](\mathbf{r}) = -
\frac{1}{2}\frac{\delta}{\delta n(\mathbf{r})}\left. \left[ \int
d\mathbf{r}\int d\mathbf{r}' \frac{\rho
(\mathbf{r},\mathbf{r}')\rho (\mathbf{r}',\mathbf{r})}{|\mathbf{r}
- \mathbf{r}'|}\right]
 \right|_{n(\mathbf{r}') = \bar \rho
(\mathbf{r}',\mathbf{r}')} .
\end{eqnarray}

By substituting $U_{KS}^{0}[\rho](\mathbf{r}) =
U_{KS-TDHF}^{0}[\rho](\mathbf{r})$ from Eq.~(\ref{tdhfexcorr})
into Eq.~(\ref{taylor}), we obtain the TDHF density-density
responses. The TDLS formalism and the associated CEO algebra were
first derived for the TDHF \cite{Chernyak,Tretiak} and this paper
extends these results to TDDFT. The difference between TDDFT and
TDHF equations is that in the former each order of the perturbed
density matrix has its own equation of motion with a different
electron-electron interaction potential, because the TDDFT
exchange-correlation potential, expanded in terms of
exchange-correlation kernels, is different in each order, while in
TDHF the electron-electron interaction is the same for all orders
\cite{Tretiak}.

\section{Acknowledgements}

The support of the National Science Foundation grant No
CHE-0132571 is gratefully acknowledged. We wish to thank
Dr.~Vladimir~Chernyak for many useful and stimulating discussions.

\appendix


\section{Projection of the density matrix into the interband subspace} \label{ap.1pr}

Since the many-electron wave function is represented at all times
by a single Slater determinant, the total density matrix
$\rho(\mathbf{r},\mathbf{r'},t)$ must be a projector. The
idempotent property  ($(\bar \rho)^{2}(\mathbf{r},\mathbf{r'}) =
\bar \rho(\mathbf{r},\mathbf{r'})$) allows us to project any
single particle matrix $\zeta$ into the interband ({\em p-h})
subspace \cite{Chernyak}
\begin{equation}\label{project}
 \zeta_{p-h}(\mathbf{r},\mathbf{r'}) = [[\zeta,\bar \rho],\bar \rho] (\mathbf{r},\mathbf{r'}).
\end{equation}
Consequently, not all elements of the density matrix are
independent \cite{Chernyak}. The idempotent property gives
\begin{equation}
  ((\bar\rho + \delta\rho(t))*(\bar\rho + \delta\rho(t)))(\mathbf{r},\mathbf{r'})=\bar\rho (\mathbf{r},\mathbf{r'}) + \delta\rho(\mathbf{r},\mathbf{r'},t).
  \label{proj}
\end{equation}
The number of degrees of freedom of $\delta\rho$ subject to the
condition Eq.~(\ref{proj}) is precisely the number
 of its particle-hole matrix elements, and $T(\xi
(\mathbf{r},\mathbf{r'},t))$ can therefore be expressed in terms
of $\xi (\mathbf{r},\mathbf{r'},t)$. Using Eq.~(\ref{proj}) and
Eq.~(\ref{decompos}) we get
\begin{equation}
  (\bar\rho + \xi (t) + T(\xi (t)))*(\bar\rho + \xi (t) + T(\xi (t)))(\mathbf{r},\mathbf{r'}) = \bar\rho (\mathbf{r},\mathbf{r'}) + \xi (\mathbf{r},\mathbf{r'},t) + T(\xi (\mathbf{r},\mathbf{r'},t)).
  \label{proj1}
\end{equation}
To simplify this expression we use the following relations, which
follow from Eq.~(\ref{project}): $\bar\rho(\mathbf{r},\mathbf{r'})
= (\bar\rho*\bar\rho)(\mathbf{r},\mathbf{r'})$, $\xi
(\mathbf{r},\mathbf{r'},t) = \bar\rho *
\xi(t)(\mathbf{r},\mathbf{r'}) + \xi(t)*\bar\rho
(\mathbf{r},\mathbf{r'})$ and $T(t)*\bar\rho
(\mathbf{r},\mathbf{r'}) = \bar\rho * T(t)
(\mathbf{r},\mathbf{r'})$. The following rule may be applied to
separate the remaining terms: product of two inter-(or two intra-)
band matrices gives an intraband matrix, whereas product of inter-
into intra- (or intra- into inter-) band matrices results in an
interband matrix. Finally, the intraband part of Eq.~(\ref{proj1})
is
\begin{equation}
 (T(\xi (t))*T(\xi (t)))(\mathbf{r},\mathbf{r'}) +  (2 \bar\rho - I)*T(\xi (t))(\mathbf{r},\mathbf{r'})+ (\xi(t)*\xi(t))(\mathbf{r},\mathbf{r'}) =
 0,
 \label{proj2}
\end{equation}
where $I$ is the unit matrix in the real space
$(I(\mathbf{r},\mathbf{r'}) = \delta (\mathbf{r} - \mathbf{r'}))$.
The formal solution of this quadratic equation, with the condition
$T(\xi(\mathbf{r},\mathbf{r'},t) = 0) = 0$ yields
 \begin{equation}
  T(\xi (\mathbf{r},\mathbf{r'},t))=
\left ( \bar \rho  - \frac{I}{2} \right )* \left ( I-
\sqrt{I-4\xi(t)*\xi(t)}\right )(\mathbf{r},\mathbf{r'}).
  \label{EQT}
 \end{equation}
Eq.~(\ref{EQTO}) is obtained by expanding Eq.~(\ref{EQT}) in
powers of $\xi$.

\section{The Second-order density response} \label{ap.2pzd}

Expanding Eq.~(\ref{eqtdhfrho}) to second order, we obtain

 \begin{eqnarray}
 \label{rho2}
 && i\frac{\partial \delta \rho_{2}(\mathbf{r},\mathbf{r}',t)}{\partial t}
  = L \delta \rho_{2}(\mathbf{r},\mathbf{r}',t) + (V_{f}[\delta\rho_{1}](\mathbf{r}, t) - V_{f}^{*}[\delta\rho_{1}](\mathbf{r'}, t))\delta \rho_{1}(\mathbf{r},\mathbf{r'},t)\nonumber \\
  &+&  (V_{g}'[\delta\rho_{1}\delta\rho_{1}](\mathbf{r}, t) - V_{g}^{'*}[\delta\rho_{1}\delta\rho_{1}](\mathbf{r'}, t))\bar \rho(\mathbf{r},\mathbf{r'})
 + (U_{1}(\mathbf{r}, t) - U_{1}^{*}(\mathbf{r'}, t))\delta \rho_{1}(\mathbf{r},\mathbf{r'},t),
  \end{eqnarray}
where
 \begin{eqnarray}
 \label{Vg1}
V_{f}[\delta\rho_{1}](\mathbf{r},t) = \int d\mathbf{r}''
f_{xc}'[\bar \rho] (\mathbf r, \mathbf r', \mathbf r'')\delta
\rho_{1}(\mathbf{r}',\mathbf{r}'',t)  ,
\end{eqnarray}
and
 \begin{eqnarray}
 \label{Vg}
V_{g}'[\delta\rho_{1}\delta\rho_{1}](\mathbf{r},t) &=&  \int
d\mathbf{r}' \int d\mathbf{r}''\int d\mathbf{r}'''\int
d\mathbf{r}'''' g_{xc}[\bar \rho] (\mathbf r, \mathbf r',
\mathbf r'',\mathbf r''',\mathbf r'''') \nonumber \\
&& \delta \rho_{1}(\mathbf{r}',\mathbf{r}'',t)\delta
\rho_{1}(\mathbf{r}''',\mathbf{r}'''',t)  ,
\end{eqnarray}
where
 \begin{eqnarray}
 \label{g''}
g_{xc}[\bar \rho] (\mathbf r, \mathbf r', \mathbf r'',\mathbf
r''',\mathbf{r}'''') = g_{xc}[\bar \rho] (\mathbf r, \mathbf r',
\mathbf{r}''')\delta(\mathbf r'- \mathbf r'')\delta(\mathbf r'''-
\mathbf r'''').
  \end{eqnarray}
Using Eq.~(\ref{eqtdhf22222}), the equation of motion for
$z_{\alpha}^{(2)}$ is
\begin{eqnarray} \label{z2}
i \frac{\partial z_{\alpha}^{(2)}(t)}{\partial t} &=&
\Omega_{\alpha} z_{\alpha}^{(2)}(t) + \sum_{\beta}
z_{\beta}^{(1)}(t) \int
 {U _{1}}(\mathbf r, t)
\rho_{-\alpha,\beta}(\mathbf r)
d\mathbf{r} \nonumber \\
&+& \sum_{\beta\gamma}V_{g(-\alpha\beta\gamma)} z_{\beta}^{(1)}(t)
z_{\gamma}^{(1)}(t), \nonumber
\\ \alpha, \beta, \gamma= \pm
1, \pm 2, \ldots,
\end{eqnarray}
where
\begin{eqnarray}
\label{vabg} V_{g(\alpha, \beta \gamma)} & = & \frac {1}{2} Tr ((I
-2{\bar \rho})*((\xi_{\beta} * \xi_{\gamma} + \xi_{\gamma}
* \xi_{\beta})
V_{g}(\xi_{\alpha}) \nonumber \\
&+& (\xi_{\alpha}* \xi_{\beta} + \xi_{\beta} * \xi_{\alpha})
V_{g}(\xi_{\gamma}) + (\xi_{\alpha}* \xi_{\gamma} + \xi_{\gamma}*
\xi_{\alpha})V_{g}(\xi_{\beta}))),
\end{eqnarray}
where
\begin{eqnarray} \label{vgpr}
V_{g}(\xi_{\alpha})(\mathbf{r})\xi_{\beta} =
V_{f}[\xi_{\alpha}](\mathbf{r})\xi_{\beta} +
V_{g}'[\xi_{\alpha}\xi_{\alpha}](\mathbf{r}) \bar \rho,
\end{eqnarray}
and
\begin{eqnarray} \label{Tr}
Tr (\zeta (\mathbf r,\mathbf r')) = \int d\mathbf r \int d\mathbf
r' \zeta (\mathbf r, \mathbf r') \delta (\mathbf r - \mathbf r') =
\int \zeta (\mathbf r,\mathbf r) d\mathbf r  = \int \zeta (\mathbf
r) d\mathbf r .
\end{eqnarray}
The solution of Eq.~(\ref{z2})  is
\begin{eqnarray} \label{z2sol3}
z^{(2)}_\alpha (t)&=& i \int_{-\infty}^t  d \tau_1 \int d\mathbf r
\int d\mathbf r' s_{\alpha} G_{\alpha} (t - \tau_1)
\Gamma^{(2)}_{\alpha} (\tau_1,\mathbf r,\mathbf r'),
\end{eqnarray}
where
\begin{eqnarray} \label{z2gamma}
\Gamma^{(2)}_{\alpha} (\tau_1,\mathbf r,\mathbf r') & = &
\sum_{\beta \gamma}
V_{g(-\alpha\beta\gamma)}\int_{-\infty}^{\tau_1} d\tau_{2}
\int_{-\infty}^{\tau_1} d\tau_{3} {U _{1}} (\mathbf r,\tau_1){U
_{1}}(\mathbf r',\tau_2)\rho_{-\beta}(\mathbf
r)\rho_{-\gamma}(\mathbf r')s_{\beta}s_{\gamma}\nonumber \\
&& G_{\beta} (\tau_1 - \tau_2)  G_{\gamma} (\tau_1 - \tau_3) +
\sum_{\beta}i{U _{1}} (\mathbf
r,\tau_1) \rho_{-\alpha \beta}(\mathbf r) \nonumber \\
&& \int_{-\infty}^{\tau_1} d\tau_{2} {U _{1}}(\mathbf r',\tau_2)
\rho_{-\beta}(\mathbf r') s_{\beta} G_{\beta} (\tau_1 - \tau_2) .
\end{eqnarray}

 The time domain second-order
density-density response function $ \chi^{(2)} (t, \tau_1,
\tau_2)$ is defined \cite{Gross}:
\begin{eqnarray} \label{rest2}
\delta\rho_{2} (\mathbf r, t) &=& \delta\rho_{2} (\mathbf r,
\mathbf r, t) = \frac{1}{2}\int_{-\infty}^t d \tau_1
\int_{-\infty}^{t} d \tau_2 \int d\mathbf{r'}\int d\mathbf{r''} {U
_{1}} (\mathbf r',\tau_1) {U _{1}} (\mathbf r'',\tau_2) \nonumber
\\
&& \chi^{(2)} (t, \tau_1, \tau_2,\mathbf r, \mathbf r', \mathbf
r'').
\end{eqnarray}
Inserting Eqs.~(\ref{zlinsol3}) and (\ref{z2sol3}) into
Eq.~(\ref{eqpolar2}) and keeping all terms up to the second order
we find that the second-order response function finally has three
contributions:
\begin{eqnarray} \label{resp2}
\chi^{(2)} (t, \tau_1, \tau_2,\mathbf r, \mathbf r', \mathbf r'')
&=& \chi_{I}^{(2)} (t, \tau_1, \tau_2,\mathbf r, \mathbf r',
\mathbf r'') + \chi_{II}^{(2)}(t, \tau_1, \tau_2,\mathbf r,
\mathbf r', \mathbf r'') \nonumber \\
&+& \chi_{III}^{(2)}(t, \tau_1, \tau_2,\mathbf r, \mathbf r',
\mathbf r''),
\end{eqnarray}
where
\begin{eqnarray} \label{resp2I}
\chi_{I}^{(2)}(t, \tau_1, \tau_2,\mathbf r, \mathbf r', \mathbf
r'') &=& 2\sum_{\alpha \beta} \rho_{-\alpha, \beta}(\mathbf
r)\rho_{\alpha}(\mathbf r') \rho_{-\beta}(\mathbf
r'') \nonumber\\
&& s_{\alpha} s_{\beta} G_{\alpha} (t - \tau_1) G_{\beta} (\tau_1 - \tau_2), \\
\label{resp2II}
\chi_{II}^{(2)}(t, \tau_1, \tau_2,\mathbf r, \mathbf r', \mathbf
r'') &=& 2i \int_{\tau_2}^{t} d\tau \sum_{\alpha \beta
\gamma}V_{g(-\alpha\beta\gamma)}\rho_{\alpha} (\mathbf
r)\rho_{-\beta}(\mathbf r') \rho_{-\gamma}(\mathbf
r'') \nonumber\\
&& s_{\alpha} s_{\beta}s_{\gamma} G_{\alpha} (t - \tau) G_{\beta} (\tau - \tau_1)G_{\gamma} (\tau - \tau_2), \\
\label{resp2III}
\chi_{III}^{(2)} (t, \tau_1, \tau_2,\mathbf r, \mathbf r', \mathbf
r'')  & = & 2 \sum_{\alpha \beta} \rho_{\alpha \beta}(\mathbf r)
\rho_{-\alpha}(\mathbf r') \rho_{-\beta}(\mathbf r'') \nonumber \\
&&  s_{\alpha} s_{\beta} G_{\alpha}(t - \tau_1) G_{\beta}(t -
\tau_2).
\end{eqnarray}

The corresponding frequency domain density-density response
function $\chi^{(2)}(-\omega_s; \omega_1, \omega_2)$ is  defined
by
\begin{eqnarray} \label{resf2}
\delta\rho_{2}  (\mathbf r, \omega_s) &=& \frac{1}{2}
\int_{-\infty}^{\infty}   \frac {d \omega_1}{2 \pi}
\int_{-\infty}^{\infty} \frac{d \omega_2}{2 \pi} \int
d\mathbf{r'}\int d\mathbf{r''} \chi^{(2)} (-\omega_s; \omega_1,
\omega_2,\mathbf r, \mathbf r',\mathbf r'') \nonumber \\
&& {U _{1}} ( \mathbf r',\omega_1) {U _{1}} (\mathbf
r'',\omega_2).
\end{eqnarray}
The relation between response functions and charge densities are
obtained by comparing Eq.~(\ref{rest2}) with Eq.~(\ref{resf2}) and
using the Fourier transform Eq.~(\ref{fourier}):
\begin{eqnarray}
\label{restf2} &&\chi^{(2)} (-\omega_s; \omega_1,\omega_2,\mathbf
r, \mathbf
r',\mathbf r'') = \nonumber \\
&&\int_{-\infty}^{\infty} dte^{i \omega_s t} \int_{-\infty}^{t} d
\tau_1 e^{-i \omega_1 \tau_1} \int_{-\infty}^t d \tau_2 e^{-i
\omega_2 \tau_2} \chi^{(2)} (t, \tau_1, \tau_2,\mathbf r, \mathbf
r',\mathbf r'').
\end{eqnarray}

The second order response function is usually denoted
\begin{eqnarray}
\label{restf2a} \chi^{(2)} (\omega_s=\omega_1+\omega_2,
\omega_1,\omega_2,\mathbf r, \mathbf r',\mathbf r'') & = &
2\pi\delta(-\omega_s+\omega_1+\omega_2) \nonumber \\
&& \chi^{(2)} (\omega_1, \omega_2,\mathbf r, \mathbf r',\mathbf
r'').
\end{eqnarray}

Using Eq.~(\ref{restf2}) we finally obtain the second-order
density-density response function which is symmetric with respect
to $\omega_1$ and $\omega_2$ permutations
\begin{eqnarray}
\chi^{(2)}(\omega_1, \omega_2,\mathbf r, \mathbf r', \mathbf r'')
&=& 2 \sum_{\alpha\beta\gamma}
\frac{V_{g(-\alpha\beta\gamma)}\rho_{\alpha}(\mathbf r)
\rho_{-\beta}(\mathbf r') \rho_{-\gamma}(\mathbf r'') s_{\alpha}
s_{\beta}}{(\Omega_{\alpha} - \omega_1
- \omega_2)(\Omega_{\beta} - \omega_1)(\Omega_{\gamma} - \omega_2)} \nonumber \\
&+& \sum_{\alpha\beta} \frac{\rho_{-\alpha \beta}(\mathbf
r)\rho_{\alpha}(\mathbf r') \rho_{-\beta}(\mathbf r'') s_{\alpha}
s_{\beta}}{(\Omega_{\alpha} - \omega_1 - \omega_2)(\Omega_{\beta}
- \omega_1)}
\nonumber \\
&+& \sum_{\alpha\beta} \frac{\rho_{-\alpha \beta}(\mathbf
r)\rho_{\alpha}(\mathbf r') \rho_{-\beta}(\mathbf r'') s_{\alpha}
s_{\beta}}{(\Omega_{\alpha} - \omega_1
- \omega_2)(\Omega_{\beta} - \omega_2)} \nonumber \\
&+& \sum_{\alpha \beta} \frac{\rho_{\alpha \beta}(\mathbf
r)\rho_{\alpha}(\mathbf r') \rho_{-\beta}(\mathbf r'') s_{\alpha}
s_{\beta}}{(s_{\alpha} \Omega_{\alpha} - \omega_1)(s_{\beta}
\Omega_{\beta} - \omega_2)}, \hspace{1em} \alpha, \beta, \gamma =
\pm 1, \pm2, \ldots. \label{pol2}
\end{eqnarray}
Here and below $\Omega_\nu$, $\nu=\alpha, \beta $ is positive
(negative) for all $\nu>0$ ($\nu<0$) following to the convention
$\Omega_{-\nu}=-\Omega_\nu$.

By setting $\omega_1$ and $\omega_2$ to zero and using the
identities $s_\nu \Omega_\nu = |\Omega_\nu|$ and
$\rho_{-\nu}=\rho_\nu$ we obtain the second order static
density-density response:
\begin{eqnarray} \label{pol2a}
\chi^{(2)}(0,\mathbf r, \mathbf r', \mathbf r'') &=& 2
\sum_{\alpha \beta \gamma}
\frac{V_{g(\alpha\beta\gamma)}\rho_{\alpha}(\mathbf r)
\rho_{\beta}(\mathbf r') \rho_{\gamma}(\mathbf
r'')}{|\Omega_{\alpha}\Omega_{\beta}\Omega_{\gamma}|} +
3\sum_{\alpha\beta} \frac{\rho_{\alpha
\beta}(\mathbf r)\rho_{\alpha}(\mathbf r') \rho_{\beta}(\mathbf r'')}{|\Omega_{\alpha}\Omega_{\beta}|}, \nonumber \\
\hspace{1em}&& \alpha, \beta, \gamma = \pm 1, \pm2, \ldots.
\end{eqnarray}

\section{The third-order density response} \label{ap.2pz3}

Expanding Eq.~(\ref{eqtdhfrho}) to third order, we obtain
 \begin{eqnarray}
 \label{rho3}
  i\frac{\partial \delta \rho_{3}(\mathbf{r},\mathbf{r}',t)}{\partial t}
  &=& L \delta \rho_{3}(\mathbf{r},\mathbf{r}',t) + (V_{f}[\delta\rho_{1}](\mathbf{r}, t) - V_{f}^{*}[\delta\rho_{1}](\mathbf{r}', t))\delta \rho_{2}(\mathbf{r},\mathbf{r'},t) \nonumber \\
  &+& (V_{f}[\delta\rho_{2}](\mathbf{r}, t) - V_{f}^{*}[\delta\rho_{2}](\mathbf{r}', t))\delta \rho_{1}(\mathbf{r},\mathbf{r'},t) \nonumber \\
  &+& (V_{g}'[\delta\rho_{1}\delta\rho_{1}](\mathbf{r}, t) - V_{g}^{'*}[\delta\rho_{1}\delta\rho_{1}](\mathbf{r}', t))\delta \rho_{1}(\mathbf{r},\mathbf{r'},t) \nonumber \\
  &+& (V_{g}'[\delta\rho_{1},\delta\rho_{2}](\mathbf{r}, t) - V_{g}^{'*}[\delta\rho_{1},\delta\rho_{2}](\mathbf{r'}, t) + V_{g}'[\delta\rho_{2},\delta\rho_{1}](\mathbf{r}, t) \nonumber \\  &-& V_{g}^{'*}[\delta\rho_{2},\delta\rho_{1}](\mathbf{r'}, t))\bar
  \rho(\mathbf{r},\mathbf{r'})
  + (V_{h}[\delta\rho_{1},\delta\rho_{1},\delta\rho_{1}](\mathbf{r}, t) \nonumber \\ &-&  V_{h}^{*}[\delta\rho_{1},\delta\rho_{1},\delta\rho_{1}](\mathbf{r'}, t)) \bar
  \rho(\mathbf{r},\mathbf{r'}) + (U_{1}(\mathbf{r}, t) - U_{1}^{*}(\mathbf{r'}, t))\delta \rho_{3}(\mathbf{r},\mathbf{r'},t),
  \end{eqnarray}
where
 \begin{eqnarray}
 \label{Vh}
V_{h}[\delta\rho_{1}\delta\rho_{1}\delta\rho_{1}](\mathbf{r},t)
&=&  \int d\mathbf{r}' \int d\mathbf{r}'' \int d\mathbf{r}'''\int
d\mathbf{r}'''' \int d\mathbf{r}'''''\int d\mathbf{r}''''''
h_{xc}[\bar \rho] (\mathbf r, \mathbf
r', \mathbf r'',\mathbf r''',\mathbf r'''',\mathbf r''''', \mathbf r'''''')  \nonumber \\
&&  \delta \rho_{1}(\mathbf{r}',\mathbf{r}'',t)\delta
\rho_{1}(\mathbf{r}''',\mathbf{r}'''',t)\delta
\rho_{1}(\mathbf{r}''''',\mathbf{r}'''''',t)  ,
\end{eqnarray}
where
 \begin{eqnarray}
 \label{h''}
h_{xc}[\bar \rho] (\mathbf r, \mathbf r', \mathbf r'',\mathbf
r''',\mathbf{r}'''',\mathbf{r}''''',\mathbf{r}'''''') =
h_{xc}[\bar \rho] (\mathbf r, \mathbf r', \mathbf r''',
\mathbf{r}''''')\delta(\mathbf r'- \mathbf r'')\delta(\mathbf
r'''- \mathbf r'''')\delta(\mathbf r'''''- \mathbf r'''''').
  \end{eqnarray}
Using Eq.~(\ref{eqtdhf22222}), the equation of motion for
$z_{\alpha}^{(3)}$ is
\begin{eqnarray} \label{z3}
i \frac{\partial z_{\alpha}^{(3)}(t)}{\partial t} &=&
\Omega_{\alpha} z_{\alpha}^{(3)}(t) + z_{\beta}^{(2)}(t)
\sum_{\beta} \int
 {U _{1}}(\mathbf r, t)
\rho_{-\alpha \beta}(\mathbf r)d\mathbf{r}
\nonumber \\
&+& \sum_{\beta \gamma} z_{\beta}^{(1)}(t) z_{\gamma}^{(1)}(t)
\int {U _{1}}(\mathbf r, t) \rho_{-\alpha \beta \gamma}(\mathbf r)
d\mathbf{r}
\nonumber \\
&+& 2\sum_{\beta \gamma}
V_{g(-\alpha\beta\gamma)}z_{\beta}^{(1)}(t) z_{\gamma}^{(2)}(t) \nonumber \\
&+& \sum_{\beta \gamma \delta} V_{h(-\alpha\beta\gamma\delta)}
z_{\beta}^{(1)}(t)
z_{\gamma}^{(1)}(t)z_{\delta}^{(1)}(t)  \nonumber \\
&& \hspace{0.5em}\alpha, \beta, \gamma, \delta = \pm 1, \pm 2,
\ldots,
\end{eqnarray}
where
\begin{eqnarray}
\label{vabgd} && V_{h(\alpha, \beta \gamma \delta)}
 = \frac{1}{6} Tr ((I -2 \bar \rho)*(\xi_{\alpha}* \xi_{\delta} +
\xi_{\delta}*\xi_{\alpha}) V_{h}((I -2\bar \rho)*(\xi_{\beta}
* \xi_{\gamma}\\ \nonumber
&+& \xi_{\gamma}* \xi_{\beta})))
+ \frac{1}{6} Tr ((I-2 \bar \rho)*(\xi_{\alpha}* \xi_{\gamma} +
\xi_{\gamma}*\xi_{\alpha})V_{h}((I -2 \bar \rho)*(\xi_{\delta}
* \xi_{\beta} \\ \nonumber &+& \xi_{\beta}* \xi_{\delta})))
+ \frac{1}{6} Tr ((I-2 \bar \rho)*(\xi_{\alpha}*\xi_{\beta} +
\xi_{\beta}*\xi_{\alpha}) V_{h}((I-2 \bar \rho)*(\xi_{\delta}*
\xi_{\alpha} + \xi_{\gamma}* \xi_{\delta}))) \\
\nonumber
&-& \frac{1}{6} Tr ((\xi_{\alpha} V_{h}(\xi_{\beta}) +
V_{h}(\xi_{\beta}) \xi_{\alpha})*(\xi_{\gamma}*
\xi_{\delta} + \xi_{\delta}* \xi_{\gamma})) \\
\nonumber
&-& \frac{1}{6} Tr ((\xi_{\alpha} V_{h}(\xi_{\gamma}) +
V_{h}(\xi_{\gamma}) \xi_{\alpha})*(\xi_{\beta}
*\xi_{\delta} + \xi_{\delta}* \xi_{\beta})) \\
&-& \frac{1}{6} Tr ((\xi_{\alpha} V_{h}(\xi_{\beta}) +
V_{h}(\xi_{\delta}) \xi_{\alpha})*(\xi_{\beta}* \xi_{\gamma} +
\xi_{\gamma}* \xi_{\beta})),
\end{eqnarray}
The solution of Eq.~(\ref{z3}) is
\begin{eqnarray} \label{z3sol3}
z^{(3)}_\alpha (t)&=& i \int_{-\infty}^t  d \tau_1 \int d\mathbf r
s_{\alpha}
 G_{\alpha} (t - \tau_1) \Gamma^{(3)}_{\alpha}
(\tau_1,\mathbf r),
\end{eqnarray}
where
\begin{eqnarray} \label{z3gamma}
\Gamma^{(3)}_{\alpha} (\tau_1,\mathbf r) & = & \sum_{\beta} {U
_{1}} (\mathbf r,\tau_1) \rho_{-\alpha
\beta}(\mathbf r)z^{(2)}_\beta (\tau_1)  \nonumber \\
&+& \sum_{\beta \gamma} {U _{1}} (\mathbf r,\tau_1) \rho_{-\alpha
\beta \gamma}(\mathbf r)z^{(1)}_\beta (\tau_1)z^{(1)}_\gamma
(\tau_1) \nonumber \\
&+& 2\sum_{\beta \gamma}
V_{g(-\alpha\beta\gamma)}z_{\beta}^{(1)}(\tau_1)
z_{\gamma}^{(2)}(\tau_1) \nonumber \\
&+& \sum_{\beta \gamma \delta}
V_{h(-\alpha\beta\gamma\delta)}z_{\beta}^{(1)}(\tau_1)
z_{\gamma}^{(1)}(\tau_1)z_{\delta}^{(1)}(\tau_1) ,
\end{eqnarray}
where $z^{(1)}(\tau_1)$ and $z^{(2)}(\tau_1)$ are given by
Eqs.(\ref{zlinsol3}) and (\ref{z2sol3}).

Time domain third-order density-density response function $
\chi^{(3)} (t, \tau_1, \ldots, \tau_3)$ is defined \cite{Gross}:
\begin{eqnarray} \label{rest3}
\delta\rho_{3} (\mathbf r, t) &=& \delta\rho_{3} (\mathbf
r,\mathbf r, t) = \frac{1}{6}\int_{-\infty}^t d \tau_1
\int_{-\infty}^{t} d \tau_2 \int_{-\infty}^{t} d \tau_3 \int
d\mathbf{r'}\int d\mathbf{r''} \int d\mathbf{r'''}{U _{1}}
(\mathbf r',\tau_1)  \nonumber \\
&& {U _{1}} (\mathbf r'',\tau_2) {U _{1}} (\mathbf r''',\tau_3)
\chi^{(3)} (t, \tau_1, \tau_2,\tau_3, \mathbf r, \mathbf r',
\mathbf r'', \mathbf r''').
\end{eqnarray}

Inserting Eqs.~(\ref{zlinsol3}) and~(\ref{z2sol3}) and
(\ref{z3sol3}) into Eq.~(\ref{eqpolar2}) and retaining all terms
to third order we obtain the final expression for the third-order
response function:

\begin{equation} \label{resp3}
\chi^{(3)} (t, \tau_1, \tau_2, \tau_3,\mathbf r, \mathbf r',
\mathbf r'',\mathbf r''') = \chi_{I}^{(3)} + \chi_{II}^{(3)} +
\chi_{III}^{(3)} + \chi_{IV}^{(3)} + \chi_{V}^{(3)} +
\chi_{VI}^{(3)} + \chi_{VII}^{(3)} + \chi_{VIII}^{(3)},
\end{equation}
where
\begin{eqnarray} \label{resp3I}
\chi_{I}^{(3)}(t, \tau_1, \tau_2, \tau_3, \mathbf r, \mathbf r',
\mathbf r'',\mathbf r''') &=& - 6i \sum_{\alpha \beta \gamma}
\rho_{-\alpha \beta}(\mathbf r) \rho_{-\beta \gamma}(\mathbf r')
\rho_{\alpha}(\mathbf r'') \nonumber \\
&& \rho_{-\gamma}(\mathbf r''') s_{\alpha} s_{\beta} s_{\gamma}
G_{\alpha} (t -\tau_1) G_{\beta} (\tau_1 - \tau_2) G_{\gamma}
(\tau_2 - \tau_3),
\\ \nonumber
\chi_{II}^{(3)}(t, \tau_1 \tau_2 \tau_3,\mathbf r, \mathbf r',
\mathbf r'',\mathbf r''') &=& 12\sum_{\alpha \beta \gamma \delta}
\rho_{-\alpha\beta}(\mathbf r) V_{g(-\beta \gamma\delta)}
\rho_{\alpha}(\mathbf r') \rho_{-\gamma}(\mathbf r'')
\rho_{-\delta}(\mathbf r''') s_{\alpha} s_{\beta} s_{\gamma}
s_{\delta}
\\ \label{resp3II}
& \times & \int_{\tau_3}^{t} d \tau G_{\alpha} (t -
\tau_1)G_{\beta} (\tau_1 - \tau) G_{\gamma} (\tau - \tau_2)
G_{\delta} (\tau - \tau_3), \\ \label{resp3III}
\chi_{III}^{(3)}(t, \tau_1, \tau_2, \tau_3, \mathbf r, \mathbf r',
\mathbf r'',\mathbf r''') &=& - 6i \sum_{\alpha \beta \gamma}
\rho_{-\alpha \beta \gamma}(\mathbf r) \rho_{\alpha}(\mathbf r')
\rho_{-\beta}(\mathbf r'') \nonumber \\
&& \rho_{-\gamma}(\mathbf r''') s_{\alpha} s_{\beta} s_{\gamma}
G_{\alpha} (t - \tau_1) G_{\beta} (\tau_1 - \tau_2) G_{\gamma}
(\tau_1 - \tau_3),
\\ \nonumber
\chi_{IV}^{(3)}(t, \tau_1, \tau_2, \tau_3, \mathbf r, \mathbf r',
\mathbf r'',\mathbf r''') &=& 12 \sum_{\alpha \beta \gamma \delta}
V_{g(-\alpha \beta \gamma)} \rho_{-\gamma
\delta}(\mathbf r) \rho_{\alpha}(\mathbf r') \rho_{-\beta}(\mathbf r'') \rho_{-\delta}(\mathbf r''') s_{\alpha} s_{\beta} s_{\gamma} s_{\delta} \\
\label{resp3IV}
& \times & \int_{\tau_3}^{t} d \tau G_{\alpha} (t - \tau)
G_{\beta} (\tau - \tau_1) G_{\gamma} (\tau - \tau_2) G_{\delta}
(\tau_2 - \tau_3), \\ \nonumber
\chi_{V}^{(3)}(t, \tau_1, \tau_2, \tau_3, \mathbf r, \mathbf r',
\mathbf r'',\mathbf r''') &=& 12 i \sum_{\alpha \beta \gamma
\delta \eta} V_{g(-\alpha \beta \gamma)} V_{g(-\gamma \delta
\eta)} \rho_{\alpha}(\mathbf
r) \rho_{-\beta}(\mathbf r') \rho_{-\delta}(\mathbf r'') \nonumber \\
&& \rho_{-\eta}(\mathbf r''')
s_{\alpha} s_{\beta} s_{\gamma} s_{\delta} s_{\eta} \times\\
\label{resp3V}
&\times& \int_{\tau_3}^t d \tau \int_{\tau_3}^{\tau} d \tau'
G_{\alpha} (t - \tau) G_{\beta} (\tau
- \tau_1) G_{\gamma} (\tau - \tau')\nonumber \\
&& G_{\delta} (\tau' - \tau_2) G_{\eta} (\tau' - \tau_3), \\
\nonumber
\chi_{VI}^{(3)}(t, \tau_1, \tau_2, \tau_3, \mathbf r, \mathbf r',
\mathbf r'',\mathbf r''') &=& 6 \sum_{\alpha \beta \gamma \delta}
V_{h(-\alpha \beta \gamma \delta)}\rho_{\alpha}(\mathbf r)
\rho_{-\beta}(\mathbf r') \rho_{-\gamma}(\mathbf r'')
\rho_{-\delta}(\mathbf r''') s_{\alpha} s_{\beta} s_{\gamma}
s_{\delta}
\\  \label{resp3VI}
& \times & \int_{\tau_3}^t d \tau G_{\alpha} (t -\tau) G_{\beta}
(\tau - \tau_1) G_{\gamma} (\tau -
\tau_2) G_{\delta}(\tau - \tau_3), \\
\label{resp3VII}
\chi_{VII}^{(3)}(t, \tau_1, \tau_2, \tau_3,\mathbf r, \mathbf r',
\mathbf r'',\mathbf r''') &=& -12i \sum_{\alpha \beta \gamma}
\rho_{\alpha\beta}(\mathbf r) \rho_{-\beta \gamma}(\mathbf r')
\rho_{-\alpha}(\mathbf r'') \nonumber \\
&& \rho_{-\gamma}(\mathbf r''') s_{\alpha} s_{\beta} s_{\gamma}
G_{\alpha} (t -
\tau_1) G_{\beta} (\tau - \tau_2) G_{\gamma} (\tau_2 - \tau_3), \\
\nonumber
\chi_{VIII}^{(3)}(t, \tau_1, \tau_2, \tau_3,\mathbf r, \mathbf r',
\mathbf r'',\mathbf r''') &=& 12 \sum_{\alpha \beta \gamma \delta}
\rho_{\alpha \beta}(\mathbf r) V_{g(-\beta \gamma \delta)}
\rho_{-\alpha}(\mathbf r') \rho_{\gamma}(\mathbf r'') \rho_{-\delta}(\mathbf r''') s_{\alpha} s_{\beta} s_{\gamma} s_{\delta}  \times\\
\label{resp3VIII}
&\times& \int_{\tau_3}^t d \tau G_{\alpha} (t - \tau_1) G_{\beta}
(t - \tau) G_{\gamma} (\tau - \tau_2) G_{\delta} (\tau - \tau_3).
\end{eqnarray}

The corresponding frequency domain density-density response
function $\chi^{(3)}(-\omega_s; \omega_1, \ldots, \omega_3)$ is
defined by
\begin{eqnarray} \label{resf3}
\delta\rho_{3} (\mathbf r, \omega_s) &=& \frac{1}{6}
\int_{-\infty}^{\infty}   \frac {d \omega_1}{2 \pi}
\int_{-\infty}^{\infty} \frac{d \omega_2}{2
\pi}\int_{-\infty}^{\infty} \frac{d \omega_3}{2 \pi} \int
d\mathbf{r'}\int d\mathbf{r''} \int d\mathbf{r'''}\chi^{(3)}
(-\omega_s; \omega_1,
\omega_2,\mathbf r, \mathbf r',\mathbf r'') \nonumber \\
&& {U _{1}} ( \mathbf r',\omega_1) {U _{1}} (\mathbf
r'',\omega_2){U _{1}} (\mathbf r''',\omega_3).
\end{eqnarray}

The relation between response functions and charge densities are
obtained by comparing Eq.~(\ref{rest3}) with Eqs.~(\ref{resf3})
and using the Fourier transform Eq.~(\ref{fourier}):
\begin{eqnarray}
\label{restf3} \chi^{(3)} (-\omega_s;
\omega_1,\omega_2,\omega_3,\mathbf r, \mathbf
r',\mathbf r'',\mathbf r''') &=& \nonumber \\
&&\int_{-\infty}^{\infty} dte^{i \omega_s t} \int_{-\infty}^{t} d
\tau_1 e^{-i \omega_1 \tau_1} \int_{-\infty}^t d \tau_2 e^{-i
\omega_2 \tau_2}\int_{-\infty}^t d \tau_3 e^{-i \omega_3 \tau_3} \nonumber \\
&& \chi^{(3)} (t, \tau_1, \tau_2,\tau_3, \mathbf r, \mathbf
r',\mathbf r'',\mathbf r''').
\end{eqnarray}

The third order response function is usually denoted
\begin{eqnarray}
\label{restf3a} \chi^{(3)} (\omega_s=\omega_1+\omega_2 + \omega_3,
\omega_1,\omega_2,\omega_3,\mathbf r, \mathbf r',\mathbf
r'',\mathbf r''') & = & 2\pi\delta(-\omega_s+\omega_1+\omega_2 +
\omega_3) \nonumber \\
&& \chi^{(3)} (\omega_1, \omega_2,\omega_3,\mathbf r, \mathbf
r',\mathbf r'',\mathbf r''').
\end{eqnarray}

Using Eqs.~(\ref{restf3}) and~(\ref{restf3a}) we obtain the
following 8-term expression for the third-order density response
function (symmetrized with respect to $\omega_1, \omega_2$, and
$\omega_3$ permutations)
\begin{equation} \label{pol3}
\chi^{(3)} (\omega_1, \omega_2, \omega_3, \mathbf r, \mathbf r',
\mathbf r'',\mathbf r''') = \sum_{\omega_1 \omega_2
\omega_3}^{perm} \left( \chi_{I}^{(3)} + \chi_{II}^{(3)} +
\chi_{III}^{(3)} + \ldots \chi_{VIII}^{(3)} \right ),
\end{equation}
where
\begin{eqnarray} \label{pol3I}
&& \chi_{I}^{(3)} = \sum_{\alpha \beta \gamma}
\frac{\rho_{-\alpha\beta}(\mathbf r) \rho_{-\beta \gamma}(\mathbf
r') \rho_{\alpha}(\mathbf r'') \rho_{-\gamma}(\mathbf r''')
s_{\alpha} s_{\beta} s_{\gamma}}{(\Omega_{\alpha} - \omega_1 -
\omega_2
- \omega_3) (\Omega_{\beta} - \omega_2 - \omega_3) (\Omega_{\gamma} - \omega_3)},\\
\label{pol3II}
&& \chi_{II}^{(3)} = \sum_{\alpha \beta \gamma \delta}
\frac{\rho_{-\alpha \beta}(\mathbf r) V_{g(-\beta \gamma \delta)}
\rho_{\alpha}(\mathbf r') \rho_{-\gamma}(\mathbf r'')
\rho_{-\delta}(\mathbf r''') s_{\alpha} s_{\beta} s_{\gamma}
s_{\delta}}{(\Omega_{\alpha} - \omega_1 - \omega_2 -
\omega_3)(\Omega_{\beta} - \omega_2
- \omega_3)(\Omega_{\gamma} - \omega_2)(\Omega_{\delta} - \omega_3)},\\
\label{pol3III}
&& \chi_{III}^{(3)} = \sum_{\alpha \beta \gamma}
\frac{\rho_{-\alpha \beta \gamma}(\mathbf r) \rho_{\alpha}(\mathbf
r') \rho_{-\beta}(\mathbf r'') \rho_{-\gamma}(\mathbf r''')
s_{\alpha} s_{\beta} s_{\gamma}}{(\Omega_{\alpha} -\omega_1 -
\omega_2
- \omega_3)(\Omega_{\beta} - \omega_2 - \omega_3)(\Omega_{\gamma} - \omega_3)},\\
\label{pol3IV}
&& \chi_{IV}^{(3)} =  \sum_{\alpha \beta \gamma \delta}
\frac{2V_{g(-\alpha \beta \gamma)} \rho_{-\gamma \delta}(\mathbf
r) \rho_{\alpha}(\mathbf r') \rho_{-\beta}(\mathbf r'')
\rho_{-\delta}(\mathbf r''') s_{\alpha} s_{\beta} s_{\gamma}
s_{\delta}}{(\Omega_{\alpha} -\omega_1 - \omega_2 -
\omega_3)(\Omega_{\beta} - \omega_1)(\Omega_{\gamma} - \omega_2 -
\omega_3)(\Omega_{\delta} - \omega_3)},\\ \label{pol3V}
&& \chi_{V}^{(3)} = \nonumber \\
&& \sum_{\alpha \beta \gamma \delta \eta} \frac{2V_{g(-\alpha\beta
\gamma)} V_{g(-\gamma \delta \eta)} \rho_{\alpha}(\mathbf r)
\rho_{-\beta}(\mathbf r') \rho_{-\delta}(\mathbf r'')
\rho_{-\eta}(\mathbf r''') s_{\alpha} s_{\beta} s_{\gamma}
s_{\delta} s_{\eta}}{(\Omega_{\alpha} -\omega_1 - \omega_2 -
\omega_3)(\Omega_{\beta} - \omega_1)(\Omega_{\gamma} -\omega_2 -
\omega_3)(\Omega_{\delta} - \omega_2)(\Omega_{\eta} - \omega_3)}\\
\label{pol3VI}
&& \chi_{VI}^{(3)} = \sum_{\alpha \beta \gamma \delta}
\frac{V_{h(-\alpha \beta \gamma \delta)} \rho_{\alpha}(\mathbf r)
\rho_{-\beta}(\mathbf r') \rho_{-\gamma}(\mathbf r'')
\rho_{-\delta}(\mathbf r''') s_{\alpha} s_{\beta} s_{\gamma}
s_{\delta}}{(\Omega_{\alpha} -\omega_1 - \omega_2 -
\omega_3)(\Omega_{\beta}
- \omega_1)(\Omega_{\gamma} - \omega_2)(\Omega_{\delta} - \omega_3)},\\
\label{pol3VII}
&& \chi_{VII}^{(3)} = \sum_{\alpha \beta \gamma}
\frac{\rho_{\alpha\beta}(\mathbf r) \rho_{-\beta \gamma}(\mathbf
r') \rho_{-\alpha}(\mathbf r'') \rho_{-\gamma}(\mathbf r''')
s_{\alpha} s_{\beta} s_{\gamma}}{(\Omega_{\alpha} - \omega_1)(
\Omega_{\beta}
- \omega_2 - \omega_3)(\Omega_{\gamma} - \omega_3)},\\
\label{pol3VIII}
&& \chi_{VIII}^{(3)} = \sum_{\alpha \beta \gamma \delta}
\frac{\rho_{\alpha \beta}(\mathbf r) V_{g(-\beta \gamma \delta)}
\rho_{-\alpha}(\mathbf r') \rho_{-\gamma}(\mathbf r'')
\rho_{-\delta}(\mathbf r''') s_{\alpha} s_{\beta} s_{\gamma}
s_{\delta}}{(\Omega_{\alpha} - \omega_1)(\Omega_{\beta} - \omega_2
- \omega_3)(\Omega_{\gamma} - \omega_2)(\Omega_{\delta} -
\omega_3)}.
\end{eqnarray}
Here $\nu = \alpha, \beta, \gamma, \delta, \eta = \pm 1, \pm 2,
\ldots $ and $\Omega_\nu$ is positive (negative) for all $\nu>0$
($\nu<0$) according to the convention $\Omega_{-\nu}=-\Omega_\nu$.
Note, that in Eq.~(\ref{pol2}) the permutations over $\omega_1$
and $\omega_2$ were written explicitly. Finally, by setting
$\omega_1$, $\omega_2$ and $\omega_3$ to zero and using identities
$s_\nu \Omega_\nu = |\Omega_\nu|$ and $\rho_{-\nu}=\rho_\nu$ we
obtain the third order static density-density response:
\begin{eqnarray}
\chi^{(3)}(0, \mathbf r, \mathbf r', \mathbf r'',\mathbf r''') & =
&6\sum_{\alpha \beta \gamma} \frac{\rho_{\alpha \beta
\gamma}(\mathbf r) \rho_{\alpha}(\mathbf r') \rho_{\beta}(\mathbf
r'') \rho_{\gamma}(\mathbf r''')}{|\Omega_{\alpha}
\Omega_{\beta}\Omega_{\gamma}|} + 6\sum_{\alpha \beta \gamma}
\frac{2\rho_{\alpha\beta}(\mathbf r) \rho_{-\beta \gamma}(\mathbf
r') \rho_{\alpha}(\mathbf r'') \rho_{\gamma}(\mathbf
r''')}{|\Omega_{\alpha} \Omega_{\beta} \Omega_{\gamma}|}\nonumber \\
&+& 6\sum_{\alpha \beta \gamma \delta} \frac{4\rho_{\alpha
\beta}(\mathbf r) V_{g(-\beta \gamma \delta)}
\rho_{\alpha}(\mathbf r') \rho_{\gamma}(\mathbf r'')
\rho_{\delta}(\mathbf r''')}{|\Omega_{\alpha}\Omega_{\beta} \Omega_{\gamma}\Omega_{\delta}|}  \nonumber \\
& + &  6\sum_{\alpha \beta \gamma \delta \eta} \frac{2V_{g(\alpha
\beta \gamma)} V_{g(-\gamma \delta \eta)} \rho_{\alpha}(\mathbf r)
\rho_{\beta}(\mathbf r') \rho_{\delta}(\mathbf r'')
\rho_{\eta}(\mathbf r''')}{|\Omega_{\alpha} \Omega_{\beta}
\Omega_{\gamma}
\Omega_{\delta} \Omega_{\eta}|} \nonumber \\
&+& 6\sum_{\alpha \beta \gamma \delta} \frac{V_{h(\alpha \beta
\gamma \delta)} \rho_{\alpha}(\mathbf r) \rho_{\beta}(\mathbf r')
\rho_{\gamma}(\mathbf r'') \rho_{\delta}(\mathbf
r''')}{|\Omega_{\alpha}\Omega_{\beta}
\Omega_{\gamma}\Omega_{\delta}|}, \nonumber \\
&& \hspace{0.5em} \alpha, \beta, \gamma, \delta, \eta = \pm 1, \pm
2, \ldots. \label{pol3a}
\end{eqnarray}

\section{The Hilbert Space (TDHS) Representation of Response Functions} \label{ap.33rd}

The ordinary ground-state Kohn-Sham equations in Hilbert space are
\cite{Kohnw:Quadoi,Gross} ($\hbar = 1$)
\begin{eqnarray} \label{eqphi}
  \widehat H_{KS}[n_{0}(\mathbf{r})](\mathbf{r}) \varphi _{j }(\mathbf{r}) &=& \varepsilon_{j} \varphi _{j }(\mathbf{r}) ;
\end{eqnarray}
\begin{eqnarray} \label{n}
  \sum_{j=1}^{K}|\varphi _{j}(\mathbf{r})|^{2} = n_{0}(\mathbf{r}) ;
\end{eqnarray}
where $n_{0}(\mathbf{r})$ is a true initial ground-state charge
density of electrons (the charge of an electron $e = 1$; $
\widehat H_{KS}[n_{0}(\mathbf{r})](\mathbf{r})$   is the Kohn-Sham
Hamiltonian Eq.(\ref{hks}) with the time-independent external
field $U_{ext}(\mathbf{r}) = U_{0}(\mathbf{r})$.

The density-density linear response
Eqs.~(\ref{rest1}),~(\ref{rest2}),~(\ref{rest3}) in the TDHS
framework can be calculated in two steps, using a perturbative
expansion in the electron-electron interaction~\cite{Gross}. The
first step is the calculation of the response function
$\chi_{s}^{(1)}(\omega,\mathbf r, \mathbf r')$ of non-interacting
particles with unperturbed density $n_{0}$ in terms of the static
unperturbed Kohn-Sham orbitals $\phi _{k}(\mathbf{r})$
\begin{eqnarray} \label{xis}
 \chi_{s}^{(1)}(\omega,\mathbf r, \mathbf r') &=& \sum_{j,k}(f_{k}-f_{j})
  \frac{\phi _{j}(\mathbf{r})\phi _{k}^{*}(\mathbf{r})\phi _{j}^{*}(\mathbf{r'})\phi _{k}(\mathbf{r'})}{\omega -(\varepsilon _{j}-\varepsilon _{k})+ i\eta}.
\end{eqnarray}
Here, ($f_{k}, f_{j}$) are the occupation numbers ($0$ or $1$) of
the KS orbitals; $\varepsilon_{j}$ and $\varepsilon_{k}$ are
corresponding energy levels of the non-interacting particles. The
summation in Eq.(\ref{xis}) extends over both occupied and
unoccupied orbitals, including the continuum states.

The linear response function for the system of interacting
particles $\chi^{(1)}(\omega,\mathbf r, \mathbf r')$  is obtained
in a second step by solving Dyson-type integral  equation:
\begin{eqnarray} \label{xi1int}
 \chi^{(1)}(\omega,\mathbf r, \mathbf r') &=& \chi_{s}^{(1)}(\omega,\mathbf r, \mathbf r') + \int d\mathbf x \int d\mathbf x'
  \chi_{s}^{(1)}(\omega,\mathbf r, \mathbf x) \nonumber \\
  && \left( \frac{e^{2}}{|\mathbf x - \mathbf x'|} + f_{xc}[n_{0}] (\omega,\mathbf x, \mathbf x') \right)
  \chi^{(1)}(\omega,\mathbf x', \mathbf r') ,
\end{eqnarray}
where $f_{xc}[n_{0}] (\omega,\mathbf r, \mathbf r')$ is a Fourier
transform with respect to time of the time-dependent
exchange-correlation kernel $f_{xc}[n_{0}] (\mathbf{r},t,
\mathbf{r'},t')$:
\begin{eqnarray} \label{fxct}
f_{xc}[n_{0}] (\mathbf{r},t, \mathbf{r'},t') = \left. \frac{\delta
U_{xc}[n](\mathbf{r},t)}{\delta n(\mathbf{r'},t')} \right|_{n_{0}}
,
\end{eqnarray}
and the exact frequency-dependent linear density response is given
by
\begin{eqnarray} \label{n1int}
 n_{1}(\omega,\mathbf r) &=& \int d\mathbf y \chi_{s}^{(1)}(\omega,\mathbf r, \mathbf y)U_{1}(\mathbf y, \omega )
 \nonumber \\
 &+&  \int d\mathbf y \int d\mathbf y'
  \chi_{s}^{(1)}(\omega,\mathbf r, \mathbf y) \left( \frac{1}{|\mathbf y - \mathbf y'|} + f_{xc}[n_{0}] (\omega,\mathbf y, \mathbf y') \right)
  n_{1}(\omega,\mathbf y') .
\end{eqnarray}
The Bethe-Salpeter-type integral equation for the second order
density-density response is represented in four-dimensional
coordinate-time space
\begin{eqnarray} \label{xi2int}
 \chi^{(2)}(x,y,y') &=& \int dz \int dz' \chi_{s}^{(2)}(x,z,z') \left. \frac{\delta
U_{KS}(z)}{\delta U_{ext}(y)}\right|_{n_{0}}\left. \frac{\delta
U_{KS}(z')}{\delta U_{ext}(y')}\right|_{n_{0}} \nonumber \\
&& + \int dz \chi_{s}^{(1)}(x,z) \int dz' \int dz''
g_{xc}(z,z',z'') \chi_{s}^{(1)}(z',y)\chi_{s}^{(1)}(z'',y')
\nonumber \\
&& + \int dz \chi_{s}^{(1)}(x,z) \int dz' (w(z,z') + f_{xc}(z,z'))
\chi^{(2)}(z',y,y') ,
\end{eqnarray}
where the time-dependent second-order exchange-correlation kernel
$g_{xc}$ is defined as:
\begin{eqnarray} \label{gxct}
g_{xc}(z,z',z'') = \left. \frac{\delta ^{2} U_{xc}(z)}{\delta
n(z') \delta n(z'')} \right|_{n_{0}} ;
\end{eqnarray}
the kernel of the (instantaneous) Coulomb interaction $w(x,x')$ is
\begin{eqnarray} \label{wxx'}
w(x,x') = \frac{e^{2}\delta(t-t')}{|\mathbf{r}-\mathbf{r'}|}.
\end{eqnarray}

 The time-dependent Kohn-Sham equations for the
second-order density response are finally obtained by inserting
Eq.(\ref{xi2int}) into Eq.(\ref{rest2}):
\begin{eqnarray} \label{n2int}
n_{2}(x) &=& \frac{1}{2}\int dz \int dz'
\chi_{s}^{(2)}(x,z,z')U_{KS}[n_{1}](z)U_{KS}[n_{1}](z') \nonumber \\
&+& \frac{1}{2} \int d z \int d z' \int dz'' \chi_{s}^{(1)}(x,z)
g_{xc}(z,z',z'')n_{1}(z')n_{1}(z'') \nonumber \\
&+& \int dz \int d z' \chi_{s}^{(1)}(x,z) (w(z,z') + f_{xc}(z,z'))
n_{2}(z') .
\end{eqnarray}
Solving Eq.(\ref{n1int}) first, allows for the subsequent solution
of the self consistent Eq.(\ref{n2int}) which is quadratic in the
(effective) perturbing potential
$U_{KS}[n_{1}(\mathbf{r},t)](\mathbf{r}, t)$.

Proceeding in a similar fashion, one can set up the Dyson-type
integral equation for the third-order density response
Eq.(\ref{rest3}):
\begin{eqnarray} \label{n3int}
n_{3}(x) &=& \frac{1}{6}\int dy \int dy' \int dy''
\chi_{s}^{(3)}(x,y,y',y'')U_{KS}[n_{1}](y)U_{KS}[n_{1}](y')U_{KS}[n_{1}](y'') \nonumber \\
&+& \frac{1}{2}\int dy \int dy' \int dz \int
dz'\chi_{s}^{(2)}(x,y,y')U_{KS}[n_{1}](y)g_{xc}(y',z,z')n_{1}(z)n_{1}(z') \nonumber \\
&+& \int dy \int dy' \int dy''
\chi_{s}^{(2)}(x,y,y')U_{KS}[n_{1}](y)(w(y',y'') + f_{xc}(y',y''))
n_{2}(y'') \nonumber \\
&+& \frac{1}{6}\int dy \int dz \int dz' \int dz''
\chi_{s}^{(1)}(x,y) h_{xc}(y,z,z',z'')n_{1}(z)n_{1}(z')n_{1}(z'') \nonumber \\
&+&  \int dy \int dy' \int dy'' \chi_{s}^{(1)}(x,y)
g_{xc}(y,y',y'')n_{1}(y')n_{2}(y'') \nonumber \\
&+& \int dy \int dy' \chi_{s}^{(1)}(x,y) (w(z,z') + f_{xc}(y,y'))
n_{3}(y') ,
\end{eqnarray}
where $h_{xc}$ is the third-order functional derivative of the
time-dependent exchange-correlation potential with respect to the
time-dependent densities
\begin{eqnarray} \label{hxct}
h_{xc}(y,z,z',z'') = \left. \frac{\delta ^{3} U_{xc}(y)}{\delta
n(z) \delta n(z') \delta n(z'')} \right|_{n_{0}} .
\end{eqnarray}



\begin{thebibliography}{99}

\bibitem{Hohenbergp:Inheg}
P. Hohenberg, W. Kohn, {\em Phys. Rev. B} {\bf 136}, 864 (1964).

\bibitem{Kohnw:Quadoi}
W. Kohn, L.~J. Sham, {\em Phys. Rev. A} {\bf 137}, 1697 (1965).

\bibitem{Par} R.~G. Parr, W. Yang, {\em Density-Functional Theory of Atoms and Molecules} (Oxford, New
  York, 1989).

\bibitem{Gross3}
E. Runge,  E. K.~U. Gross, {\em Phys. Rev. Lett.} {\bf 52}, 997
(1984).

\bibitem{Gross2}
 E. K.~U. Gross, W. Kohn, {\em Phys. Rev. Lett.} {\bf 55}, 2850
(1985).


\bibitem{Beckead:Deneaw}
A.~D. Becke, {\em Phys. Rev. A} {\bf 38}, 3098 (1988).

\bibitem{trickey2}
  S.~B. Trickey, {\em Adv. Quantum Chem.}, {\bf 21} (1990).

\bibitem{Popleja:Kohdtw}
J.~A. Pople,  P. M.~W. Gill, B.~G. Johnson,  {\em Chem. Phys.
Lett.} {\bf 199},  557 (1992).

\bibitem{casida}
M.~E. Casida,   in {\em Recent Advances in Density-Functional
Methods}, Vol.~3
  of {\em Part I}, edited by D.~A. Chong  (World Scientific, Singapore, 1995).

\bibitem{burke} K. Burke,  J.~P. Perdew, M. Levy, in {\em Modern Density
Functional Theory: A Tool for Chemistry},
            edited by J.~M. Seminario   and  P. Politzer (Elsevier, Amsterdam, 1995).


\bibitem{Gross} E. K.~U. Gross, J.~F. Dobson, M. Petersilka, in {\em Density Functional
  Theory}, edited by  R.~F. Nalewajski, Vol.~181, (Springer, Berlin, 1996).

\bibitem{Gross1} M. Petersilka, U.~J. Gossmann, E. K.~U. Gross, {\em Phys. Rev. Lett.} {\bf 76}, 1212 (1996).

\bibitem{JamorskiC:Dynpes}
C. Jamorski, M.~E. Casida, ;  D.~R. Salahub, {\em J. Chem. Phys.}
  {\bf 104},  5134 (1996).


\bibitem{Onida} G. Onida, L. Reining, A. Rubio, {\em Reviews of Modern Physics} {\bf 74},  601 (2002).


\bibitem{Coleman} A. J. Coleman and V. I. Yukalov, \emph{Reduced Density
Matrices} (Springer-Verlag Berlin Heidelberg 2000).

\bibitem{Mukamel} S. Mukamel, {\em Principles of Nonlinear Optical Spectroscopy} (Oxford, New
  York, 1995).



\bibitem{Takahashia:Anhmns}
A. Takahashi, S. Mukamel, {\em J. Chem. Phys.} {\bf 100}, 2366
(1994).


\bibitem{Chernyak} V. Chernyak, S. Mukamel, {\em J. Chem. Phys.} {\bf 103},  7640 (1995);
ibid V. Chernyak, S. Mukamel, {\em J. Chem. Phys.} {\bf 104},  444
(1996).


\bibitem{Tretiak1} S. Tretiak, V. Chernyak, S. Mukamel, {\em J. Am. Chem. Soc.}   {\bf 119},  11408 (1997).

\bibitem{Tretiak} S. Tretiak, S. Mukamel, {\em Chemical Reviews} {\bf 102},  3171 (2002).



\bibitem{Chernyak2} V. Chernyak, S. Mukamel, {\em J. Chem. Phys.} {\bf 112},  3572 (2000).

\bibitem{White} S.~R. White, Phys. Rev. Lett. {\bf 69}, 2863 (1992).


\bibitem{GrossKohn} E. K.~U. Gross and W. Kohn, {\em Adv. Quantum Chem.} {\bf 21}, 255 (1990).

\bibitem{thoulessbook}
D.~J. Thouless, {\em The Quantum Mechanics of Many-Body Systems}
(Academic
  Press, New York, 1972).

\bibitem{ring}
P. Ring, P. Schuck, P. {\em The Nuclear Many-Body Problem}
(Springer-Verlag, New
  York, 1980).

\bibitem{blaizot}
J.-P. Blaizot, G. Ripka, {\em Quantum Theory of Finite Systems}
(The MIT
  Press, Cambridge Massachusetts, 1986).

\bibitem{Bloembergen} N. Bloembergen, {\em Nonlinear Optics}
(Benjamin, New York, 1965).

\end{thebibliography}
\end{document}